\documentclass[journal]{IEEEtran}
\ifCLASSINFOpdf
\else
\fi
\usepackage{amssymb}
\usepackage{amsmath}

\usepackage{graphicx}
\usepackage{pifont}

\usepackage{array}
\usepackage{tikz}
\usepackage{booktabs}

\usepackage{xcolor}
\usepackage[colorlinks=true, citecolor=blue]{hyperref}

\definecolor{mygreen}{RGB}{0,150,0}

\definecolor{emptycirc}{RGB}{255,255,255}  
\definecolor{halfcirc}{RGB}{0,0,0}         
\definecolor{fullcirc}{RGB}{169,169,169}    

\newcommand{\customcirc}[1]{\tikz[baseline=(char.base)]{\node[shape=circle,draw,inner sep=2pt,fill=#1] (char) {};}}
\usepackage{algorithm}
\usepackage{algorithmic}
\usepackage{hyperref}
\usepackage{tikz}
 
\usepackage{graphicx}
\usepackage{booktabs}
\usepackage{xspace}

\usepackage{multirow}
\usepackage{makecell}

\usepackage{ mathrsfs }
\usepackage{graphicx, adjustbox}
\usepackage{array}
\usepackage{setspace}
\usepackage{enumitem}
\usepackage{breqn}
\usepackage{bm}
\usepackage{ dsfont }
\usepackage{ upgreek }
\usepackage{textcomp}
\usepackage{multirow}
\usepackage{algorithm}
\usepackage{algorithmic}
\usepackage{caption}
\usepackage{pgfplots}
\usepackage{tkz-euclide}
\usepackage{cite}
\usepackage{subcaption}
\usepackage{tikz}{\tiny }
\usepackage{balance}
\usepackage{ tipa }

\usepackage{fancyhdr}
\usepackage{ upgreek }
\usepackage{soul}
\usepackage{breqn}
\usepackage[english]{babel}
\usepackage{braket} 

\addto\captionsenglish{}
\allowdisplaybreaks

\usepackage{amsthm}

\usepackage{fancyhdr}
\usepackage{ upgreek }
\usepackage{soul}
\makeatletter
\newcommand{\vast}{\bBigg@{4}}

\newcommand{\Vast}{\bBigg@{5}}
\begin{document}

\title{Scalable Malware Family Classification Using Quantum Kernel–Based Machine Learning}
\author{Ratun Rahman, Hassan Jalil Hadi, Christopher Gabriel Pedraza Pohlenz, and Ali Shoker%

\thanks{Ratun Rahman is with the Department of Electrical and Computer Engineering, University of Alabama in Huntsville, Huntsville, AL 35899, USA. Email: rr0110@uah.edu.}

\thanks{Hassan Jalil Hadi*, Christopher Gabriel Pedraza Pohlenz, and Ali Shoker are with the CyberSAR Research Center, King Abdullah University of Science and Technology (KAUST), Thuwal, Jeddah 23955-6900, Saudi Arabia. Emails: hassan.hadi@kaust.edu.sa, christopher.pedrazapohlenz@kaust.edu.sa, ali.shoker@kaust.edu.sa.}

\thanks{Corresponding author: Hassan Jalil Hadi, email: hassan.hadi@kaust.edu.sa.}
\thanks{This work has been submitted to the IEEE Transactions on Dependable and Secure Computing for possible publication.}
}



\maketitle

\begin{abstract}
The classification of malware families is a key challenge in cybersecurity, which enables threat attribution, analysis of attack operations, and the formulation of effective defense strategies. Emerging malware samples are becoming increasingly structurally similar and obfuscated, making accurate multiclass classification challenging for traditional machine learning models, especially when deployed at scale. In this research, we propose a scalable Quantum Kernel-based Machine Learning (QKML) framework for malware family classification that addresses both accuracy and efficiency constraints. The proposed framework extracts structural features from executable files and uses a supervised Linear Discriminant Analysis (LDA) projection to generate a compact, class-aware representation well suited for quantum processing. The nonlinear relationships among malware families are captured using a fidelity-based quantum kernel built from parameterized quantum circuits. We use the Nystr\"{o}m approximation method to obtain a low-rank approximation of the quantum kernel, which enables effective multiclass classification via ridge regression and enables learning from all available training samples without incurring the quadratic computational cost of kernel matrix construction. The proposed model achieves strong classification performance, with 80.88\% accuracy, outperforming classical machine learning baselines under identical feature and data splits, according to experimental evaluation on a large-scale malware dataset that includes 18,836 samples across 23 malware families. These findings suggest that scalable quantum-kernel-based machine learning can offer measurable performance advantages for real-world malware family classification tasks.
\end{abstract}

\maketitle



\section{Introduction}\label{sec:introduction}

Malware family classification is a fundamental task in cybersecurity because it enables threat attribution, malware lineage analysis, and the detection of coordinated attack campaigns \cite{sharma2025characterization, chen2025multi, wu2025malscan}. Security analysts can more effectively understand adversarial tactics and develop efficient detection and mitigation strategies by classifying malicious software into families based on shared structural and behavioral traits \cite{gebrehans2025generative, ahmad2025geaad}. However, the complexity of this task has substantially increased due to the rapid proliferation of malware variants, which is driven by widespread obfuscation \cite{sharma2025mosdroid}, packing \cite{zheng2025gupacker}, and automated malware generation techniques \cite{manikandan2025next}. Modern malware samples often exhibit high inter-family similarity while also producing large numbers of polymorphic variants, making accurate and scalable family-level malware classification increasingly difficult for existing security analytics pipelines \cite{hasan2025enhancing}. Numerous classical machine learning and deep learning techniques, such as linear classifiers \cite{udayakumar2018malware}, tree-based models \cite{yeboah2020classification}, kernel methods \cite{singh2024malware}, and neural networks \cite{kalash2018malware}, have been used to classify malware families to address these challenges. Although these methods have shown promising results in controlled environments, their effectiveness frequently declines in large-scale, highly multiclass settings. Handcrafted features, in particular, may fail to capture subtle nonlinear relationships among malware families, whereas deep learning models typically require extensive labeled data and computational resources, leading to reduced interpretability in security-sensitive applications. Furthermore, the presence of pronounced class imbalance and structural similarity among malware families can limit the generalization capability of classical models \cite{pachhala2021comprehensive, ronen2018microsoft, aboaoja2022malware}, highlighting the need for learning frameworks that can better capture complex inter-family relationships without imposing prohibitively high computational costs.

\begin{figure*}
    \centering
    \includegraphics[width=0.99\linewidth]{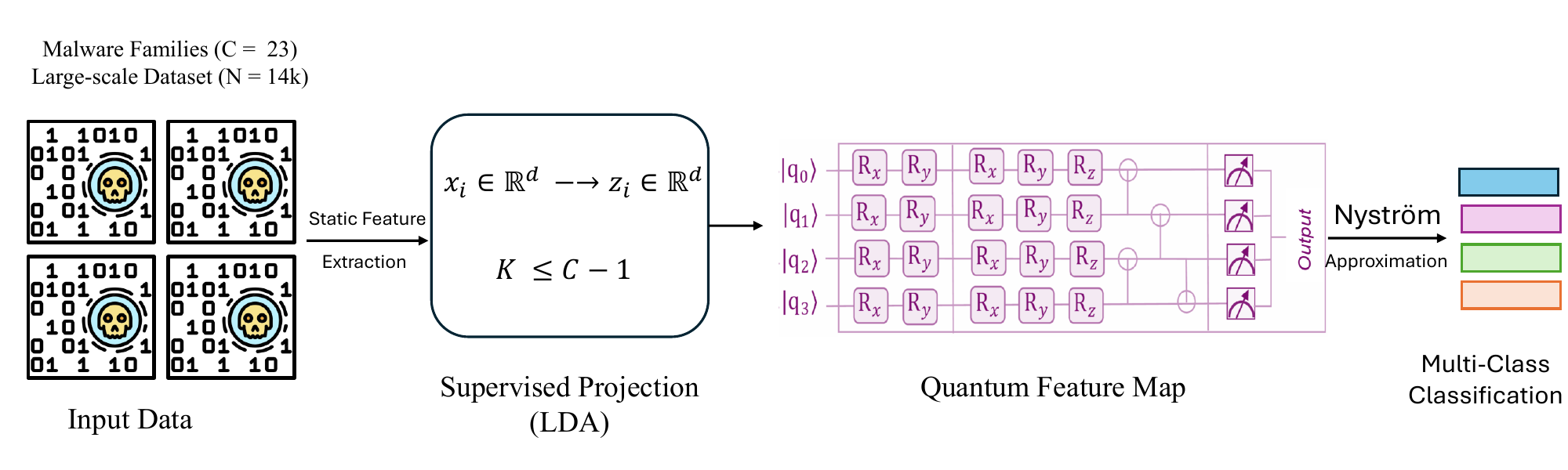}
    \caption{An overview of the proposed scalable QML framework for malware family classification. A large-scale malware dataset with $N \approx 14$k samples from $C=23$ malware families is first used to extract static features. Supervised Linear Discriminant Analysis (LDA) projects the resulting feature vectors into a low-dimensional, label-aware subspace whose dimensionality is bounded by $K \leq C-1$. A fidelity-based quantum kernel is then constructed by encoding the projected features using a parameterized quantum feature map. After applying the Nystr\"{o}m approximation to the quantum kernel to enable scalable learning on large datasets, a classical ridge regression model performs multiclass classification.}
    \label{fig:placeholder}
\end{figure*}

The capacity of quantum feature maps to embed classical data into high-dimensional Hilbert spaces has made Quantum Machine Learning (QML) a promising paradigm for improving cybersecurity pattern recognition tasks in recent years \cite{faruk2022review, paul2025integration}. Specifically, quantum kernel techniques use parameterized quantum circuits to construct similarity metrics that may offer richer representations than classical kernel functions by capturing complex nonlinear relationships among data samples. Due to code reuse, packing, and obfuscation techniques, various malware families frequently share overlapping structural characteristics, making such properties particularly relevant for malware family classification \cite{el2024quantum}. Despite these advantages, the scope and applicability of current quantum-based malware detection and classification research remain limited. Most previous studies focus on binary classification problems or a small number of malware classes, limiting their applicability to practical threat analysis where multiclass family-level attribution is essential \cite{barrue2023quantum, biswas2024sensor}. Furthermore, many studies evaluate quantum models on small or synthetic datasets because the construction of full quantum kernel matrices scales quadratically with the number of samples, making direct application to large malware repositories computationally impractical \cite{mercaldo2022towards}. As a result, subsampling is often used, which can introduce bias and degrade classification performance. In addition, existing approaches frequently overlook the challenge of integrating quantum kernels with scalable classical learning pipelines, raising questions about how quantum models can be applied effectively to large-scale, real-world malware datasets \cite{bikku2024enhancing, ciaramella2022introducing}. In multiclass cybersecurity settings, these constraints highlight the need for scalable QML frameworks that can fully utilize available data while preserving computational feasibility.

To address this gap between the expressive potential of QML and the scalability requirements of practical malware analysis, this research proposes a scalable QKML framework for multiclass malware family classification. The core idea is to integrate quantum kernel modeling, efficient kernel approximation, and supervised feature projection into a single learning pipeline. The proposed framework is shown in Fig.~\ref{fig:placeholder}. We summarize our key contributions as follows. 

\begin{itemize}
    \item Existing quantum kernel-based approaches have difficulty scaling to large malware datasets due to the quadratic cost of kernel matrix creation and the reliance on subsampling, which causes bias. In order to address this, we suggest a scalable quantum kernel machine learning (QKML) framework that combines a fidelity-based quantum kernel with Linear Discriminant Analysis (LDA). This allows for class-aware dimensionality reduction and enhances the discriminative power of quantum feature representations.
    \item The practical application of quantum kernel assessment is limited by its high processing cost. In order to overcome this difficulty, we build a low-rank representation of the quantum kernel from a limited number of landmark samples using the  Nystr\"{o}m approximation. This allows training on the entire dataset without subsampling while significantly reducing the number of required quantum fidelity evaluations.
    \item We curate and preprocess an extensive malware dataset of 18,836 samples from 23 malware families in order to assess the usefulness of QKML in real-world scenarios. We use ridge regression on the approximated quantum feature space to create a multiclass classification pipeline after extracting structural features from executable files.
    \item We show through comprehensive experiments that the proposed framework regularly outperforms standard machine learning baselines under the same settings, achieving over 80.88\% classification accuracy. These findings demonstrate the promise of scalable quantum kernel techniques for practical cybersecurity uses.
\end{itemize}

\section{Related Work}
\subsection{Malware Datasets}
Understanding malware behavior remains a fundamental challenge in modern cybersecurity research. Traditional malware detection approaches often rely on static signatures, handcrafted features, or API-call sequences extracted through dynamic analysis. While API-call traces obtained via sandbox execution are widely used to approximate runtime behavior~\cite{b23}, such representations are often limited in scalability, reproducibility, and structural richness, particularly when advanced obfuscation or packing techniques are involved.
\begin{table*}[h]
\captionsetup{justification=centering}
\caption{Availability comparison of related PE malware datasets with emphasis on multi-view representations and quantum-learning readiness. 
\protect\customcirc{emptycirc} = ``Not Available" \protect\customcirc{halfcirc} = ``Partially Available"  \protect\customcirc{fullcirc} = ``Available"}
\label{tab:updated_dataset_comparison}
\scriptsize
\renewcommand{\arraystretch}{1.4}

\begin{tabular*}{\linewidth}{@{\extracolsep{\fill}}cccccccccccc}
\toprule
\textbf{Dataset} & 
\textbf{FCG} & 
\textbf{Decompiled} & 
\textbf{Multi-View} & 
\textbf{Static} & 
\textbf{Family} & 
\textbf{Q-Feature} & 
\textbf{Quantum-Ready} &
\textbf{Samples} & 
\textbf{Benign} & 
\textbf{Malware} & 
\textbf{Collection Period} \\
\midrule

Malicia \cite{b16} & 
\customcirc{emptycirc} & 
\customcirc{emptycirc} & 
\customcirc{emptycirc} & 
\customcirc{halfcirc} & 
\customcirc{fullcirc} &
\customcirc{emptycirc} &
\customcirc{emptycirc} &
11,363 & 0 & 11,363 & 
2012--2013 \\

BIG-2015 \cite{b18} & 
\customcirc{emptycirc} & 
\customcirc{emptycirc} & 
\customcirc{halfcirc} & 
\customcirc{halfcirc} & 
\customcirc{fullcirc} &
\customcirc{emptycirc} &
\customcirc{emptycirc} &
10,643 & 0 & 10,643 & 
Prior to 2015 \\

EMBER \cite{b21} & 
\customcirc{emptycirc} & 
\customcirc{emptycirc} & 
\customcirc{emptycirc} & 
\customcirc{fullcirc} & 
\customcirc{emptycirc} &
\customcirc{emptycirc} &
\customcirc{emptycirc} &
2,050,000 & 750,000 & 800,000 & 
2017--2018 \\

BODMAS \cite{b22} & 
\customcirc{emptycirc} & 
\customcirc{emptycirc} & 
\customcirc{emptycirc} & 
\customcirc{fullcirc} & 
\customcirc{halfcirc} &
\customcirc{emptycirc} &
\customcirc{emptycirc} &
134,434 & 77,142 & 54,756 & 
2019--2020 \\

SOReL\_20M \cite{b20} & 
\customcirc{emptycirc} & 
\customcirc{emptycirc} & 
\customcirc{emptycirc} & 
\customcirc{fullcirc} & 
\customcirc{emptycirc} &
\customcirc{emptycirc} &
\customcirc{emptycirc} &
9,962,820 & 9,762,177 & 9,962,820 & 
2018--2019 \\

SBAN \cite{jelodar2025sban} & 
\customcirc{emptycirc} & 
\customcirc{fullcirc} & 
\customcirc{fullcirc} & 
\customcirc{emptycirc} & 
\customcirc{emptycirc} &
\customcirc{halfcirc} &
\customcirc{halfcirc} &
3,300,000 & N/A & N/A & 
2023-2024 \\

\textbf{QLCD (Ours)} & 
\customcirc{fullcirc} & 
\customcirc{fullcirc} & 
\customcirc{fullcirc} & 
\customcirc{fullcirc} & 
\customcirc{fullcirc} &
\customcirc{fullcirc} &
\customcirc{fullcirc} &
\textbf{19,000} & 
\textbf{982} & 
\textbf{18K} & 
\textbf{2023--2026} \\

\bottomrule
\end{tabular*}
\end{table*}
Over the past decade, several benchmark datasets have been introduced to support machine learning-based malware detection. The Microsoft BIG-2015 Challenge dataset~\cite{b18}, released in 2015, contains approximately 10,000 malware samples across nine families. However, the dataset provides binary content primarily in hexadecimal form without complete PE header information, decompiled code, or benign samples, limiting its applicability for structural or multi-view analysis.

EMBER~\cite{b21}, initially introduced in 2017 and updated in 2018, significantly expanded dataset scale by incorporating over two million PE files represented as engineered static feature vectors. While EMBER includes benign samples and supports classical machine learning research, it does not provide raw multi-modal program representations such as decompiled source code or function call graphs \cite{hadi2025fcg}. Similarly, the BODMAS dataset~\cite{b22} emphasizes temporal analysis of PE malware but focuses primarily on feature-based representations rather than structural program abstractions.

In 2020, SOREL-20M~\cite{b20} introduced a large-scale benchmark comprising nearly ten million samples. Although it substantially increased dataset size and temporal diversity, its representations remain feature-centric and do not include aligned multi-view program abstractions suitable for semantic reasoning or graph-based learning. Other datasets, such as UCSB-Packed~\cite{b19}, concentrate on packed malware and static classifier robustness but lack comprehensive structural representations and family-level multi-view alignment. More recently, SBAN~\cite{jelodar2025sban} introduced a large-scale multi-dimensional dataset integrating binary code, assembly instructions, source code, and natural language descriptions into a unified benchmark. SBAN comprises over 3 million samples across four aligned representation layers, enabling cross-modal learning and LLM pre-training for software code mining. While SBAN advances multimodal code representation for language modeling and malware analysis, it is primarily designed for classical LLM pre-training and semantic software mining rather than structured quantum feature encoding or hybrid quantum-classical learning pipelines.

Despite these advancements, existing datasets predominantly target classical machine learning pipelines and feature-based detection tasks. None are explicitly designed to support structured multi-view alignment across binaries, assembly, decompiled code, and graph representations. Moreover, they do not provide feature normalization or encoding schemes tailored for hybrid quantum-classical learning architectures. Consequently, there remains a gap in datasets that simultaneously enable semantic code understanding, structural reasoning, malware family attribution, and quantum-ready feature construction, motivating the development of QLCD. The comparison result with different malware datasets is shown in Table~\ref{tab:updated_dataset_comparison}.

\subsection{Classical Malware Family Classification}
The classification of malware families has been extensively investigated using traditional machine learning methods, especially in the static analysis of Portable Executable (PE) files. To train traditional classifiers, early methods focused on extracting handcrafted features such as byte-level statistics \cite{yuan2020byte}, opcode frequencies \cite{soni2022opcode}, section entropy \cite{zhu2022few}, and structural metadata \cite{li2024meta}. Large malware corpora benefit from the scalability and low computational overhead of linear models, such as logistic regression \cite{pascanu2015malware} and linear support vector machines \cite{udayakumar2018malware}.

Several studies have used instance-based classifiers and kernel-based techniques to capture nonlinear relationships among malware samples \cite{park2025malcl}. By utilizing similarity in feature space, support vector machines \cite{gupta2022malware} with radial basis function kernels and $k$-nearest neighbors \cite{al2025enhancing} have been shown to improve classification accuracy. However, as the number of malware families increases, these approaches often suffer from increased computational complexity and sensitivity to feature scaling and class imbalance. Probabilistic models such as Naive Bayes classifiers \cite{pachhala2021comprehensive} have also been investigated for malware family classification. These models offer fast training and inference, but they usually perform poorly because of strong independence assumptions that are rarely satisfied in real-world malware data.

Even with significant advances \cite{zhang2025imcmk, yu2025semantic, li2025malmixer, rose2025malware}, modeling complex structural relationships resulting from code reuse, polymorphism, and obfuscation techniques remains a challenge for traditional malware family classification methods. The expressivity of classical models becomes a limiting factor as malware families show overlapping feature distributions and share increasingly common components. These limitations have prompted recent interest in more expressive similarity-based learning frameworks, such as advanced kernel techniques and quantum-inspired methods \cite{sharma2025characterization}. These frameworks seek to better capture subtle relationships among malware samples while preserving scalability.

\subsection{QML for Cybersecurity}
In recent years, interest in QML has increased as an emerging paradigm for improving data-driven security analytics. Learning-based systems are frequently used in cybersecurity for tasks including malware detection \cite{el2024quantum}, intrusion detection \cite{kim2024quantum}, and anomaly analysis \cite{rahman2025toward}, where adversarial behaviors, high-dimensional data, and nonlinear correlations pose persistent challenges. In these contexts, QML seeks to complement traditional learning techniques by utilizing quantum mechanical principles to create expressive feature representations and similarity metrics. In these works, classical features are usually encoded into quantum states using parameterized quantum circuits, and similarity evaluation or classification is then based on measurements. These preliminary findings motivate further research into the potential of quantum feature maps for security applications, as they indicate that such methods can capture complex correlations that are challenging to represent with traditional linear or shallow nonlinear classifiers \cite{faruk2022review, paul2025integration}.

However, there are still several key barriers preventing QML from being widely used in cybersecurity. The majority of current research focuses on binary classification tasks, small-scale datasets, or extremely simplified feature representations that do not accurately capture the complexity of real-world cyberthreats \cite{barrue2023quantum, biswas2024sensor, faruk2022review}. Furthermore, many QML techniques rely on computationally costly kernel evaluations that scale quadratically with the size of the dataset or assume idealized quantum hardware, which makes them unsuitable for large operational datasets that are frequently used in network monitoring and malware analysis \cite{bikku2024enhancing, ciaramella2022introducing}. In cybersecurity, where scalability, robustness, and interpretability are crucial requirements, these limitations are particularly significant.

QML frameworks that specifically address scalability and integration with traditional security pipelines are therefore becoming increasingly important \cite{vyas2025comparison, kar2025unified}. This means integrating quantum-enhanced similarity learning with traditional preprocessing, dimensionality reduction, and effective approximation methods for cybersecurity tasks involving malware family classification. Beyond proof-of-concept demonstrations, evaluating the full potential of QML in real-world cybersecurity scenarios requires the development of such hybrid and scalable approaches.

\subsection{Kernel-Based and Quantum Learning for Malware Analysis}
Because kernel-based learning can model nonlinear relationships among samples without explicitly increasing feature dimensionality, it has become an important extension of traditional malware classification frameworks. Using similarity metrics derived from static features, behavioral traces, or graph-based representations, kernel methods have been used in the security literature for malware detection and family classification. Kernel techniques have shown improved discrimination capability in situations involving polymorphism and feature overlap across malware families by implicitly mapping samples into high-dimensional feature spaces \cite{zhang2024ranker}. However, the scalability of kernel-based models to large malware repositories is often limited due to their quadratic computational complexity with respect to the number of training samples.

Quantum kernels have emerged as a promising alternative for creating expressive similarity measures in recent developments in QML. Quantum kernels use parameterized quantum circuits to embed classical data into quantum Hilbert spaces, where state fidelity or similar metrics are used to quantify similarity. Several investigations into quantum kernel techniques for classification tasks have shown improved performance on specific benchmarks and synthetic datasets. Early research indicates that quantum feature maps may capture intricate correlations in security applications that are challenging to model using classical kernels alone \cite{bikku2024enhancing, passo2025supply}.

Despite their theoretical potential, current quantum learning techniques have serious drawbacks that restrict their use in real-world malware analysis. The majority of earlier research relies on limited feature representations, binary classification settings, or small-scale datasets that do not accurately represent the complexity of operational malware ecosystems. Furthermore, straightforward implementations are not feasible for large malware corpora due to the quadratically increasing computational cost of evaluating full quantum kernels. Additionally, many studies overlook scalability issues and integration with traditional preprocessing pipelines that are frequently employed in security analytics in favor of idealized quantum hardware or simulation settings.

These drawbacks draw attention to a crucial disconnect between the expressive potential of quantum kernel techniques and their usefulness in large-scale malware family classification. Scalable kernel approximation methods, careful integration with traditional dimensionality reduction, and evaluation on realistic, multiclass malware datasets are necessary to close this gap. To close this gap and enable the practical implementation of quantum-enhanced learning within real-world malware analysis workflows, the current work combines supervised feature projection with scalable Nystr\"{o}m-approximated quantum kernels.

\subsection{Summary of Limitations and Research Gap}
Several significant limitations that impede practical deployment in real-world cybersecurity settings are highlighted by the current body of work on malware family classification and quantum-enhanced learning. Despite being scalable and well understood, traditional machine learning techniques frequently lack the expressivity required to capture intricate structural relationships among malware families that result from code reuse, polymorphism, and obfuscation. Although kernel-based extensions enhance discrimination capabilities, their applicability to large malware repositories is limited due to their quadratic computational complexity.

Recent work in QML presents promising methods for creating expressive similarity measures using quantum kernels and parameterized quantum feature maps. Nevertheless, previous QML-based cybersecurity research has mostly focused on binary classification problems, small-scale datasets, or simplified feature representations. Furthermore, as dataset sizes grow, the majority of current quantum kernel implementations require full kernel evaluation, which results in prohibitively high computational costs. The assumptions of idealized quantum hardware and the lack of integration with traditional preprocessing pipelines frequently employed in operational security analytics further exacerbate these limitations.

Consequently, there is a clear research gap between the theoretical expressivity of quantum kernel methods and their usefulness for large-scale, multiclass malware family classification. A learning framework that integrates scalable kernel approximation, supervised dimensionality reduction, expressive quantum similarity modeling, and evaluation on real-world malware datasets is needed to close this gap. By introducing a scalable quantum kernel learning pipeline that uses real-world malware features, employs the Nystr\"{o}m approximation for efficiency, and systematically evaluates performance in a multiclass cybersecurity setting, the current work aims to address this gap.

\section{Methodology}
The proposed QKML framework for scalable malware family classification is presented in this section. Let \(
\mathcal{D} = \{(\mathbf{x}_i, y_i)\}_{i=1}^{N} \) represent the malware dataset. The feature vector extracted from the $i$-th malware sample is denoted by $\mathbf{x}_i \in \mathbb{R}^d$, and the corresponding malware family label among $C$ classes is denoted by $y_i \in \{1,2,\dots,C\}$. For large-scale datasets, the objective is to learn a multiclass classifier that can accurately predict $y_i$ from $\mathbf{x}_i$ while maintaining computational efficiency.

The four primary components of the proposed framework are:  
\begin{itemize}
    \item Feature representation and dataset construction;
    \item Supervised feature projection using linear discriminant analysis;
    \item Quantum kernel construction using parameterized quantum circuits; and
    \item Scalable kernel approximation and multiclass classification.
\end{itemize}
The following subsections describe each component.

\subsection{Dataset Construction and Feature Representation}
We collect executable files labeled with malware family information to construct a large-scale malware dataset. Static analysis is used to extract structural features from the Portable Executable (PE) format for each sample. These features are designed to capture commonalities introduced by code reuse, packing, and obfuscation strategies used by various malware families.

Let \(\mathbf{x}_i = [x_{i1}, x_{i2}, \dots, x_{id}]^\top \in \mathbb{R}^d\) represent the raw feature vector associated with the malware sample indexed by $i$. In addition to categorical attributes such as machine type and compiler or packer information, the extracted features also include numerical attributes, including global file entropy, file size, number of sections, statistical summaries of section entropies, and ratios between virtual and raw section sizes. One-hot encoding is used to convert categorical features into numerical form, creating a unified feature space.

Z-score normalization is applied to standardize all numerical features to ensure numerical stability and consistent scaling across dimensions. In particular, each feature dimension is transformed as
\begin{equation}
\tilde{x}_{ij} = \frac{x_{ij} - \mu_j}{\sigma_j},
\end{equation}
where $\mu_j$ and $\sigma_j$ represent the mean and standard deviation of the $j$-th feature computed over the training set. Before normalization, median imputation is used to handle missing feature values.

Malware families with insufficient representation are excluded to enable reliable multiclass learning. Let $N_c$ be the number of samples in class $c$; only classes satisfying
\begin{equation}
N_c \geq N_{\min}
\end{equation}
are retained, where $N_{\min}$ denotes a predefined minimum sample threshold. Following this filtering step, the final dataset consists of $N$ samples from $C$ malware families.

Training and test sets, denoted by $\mathcal{D}_{\mathrm{train}}$ and $\mathcal{D}_{\mathrm{test}}$, respectively, are obtained by randomly splitting the dataset. $\mathcal{D}_{\mathrm{train}}$ is used exclusively to learn all subsequent model components, including feature projection, quantum kernel construction, and classifier training, while $\mathcal{D}_{\mathrm{test}}$ is reserved for performance evaluation. This experimental procedure allows the proposed framework to fully utilize the available training data while ensuring a fair evaluation of generalization performance.

\subsection{Supervised Feature Projection via Linear Discriminant Analysis}
Due to the limited number of available qubits and the exponential growth in quantum circuit complexity with increasing input dimension, it is not feasible to directly embed high-dimensional malware features into a quantum feature space. As a supervised dimensionality reduction step prior to quantum encoding, we employ Linear Discriminant Analysis (LDA) to address this issue while preserving class-discriminative information.

Let $\mathcal{D}_{\mathrm{train}} = \{(\mathbf{x}_i, y_i)\}_{i=1}^{N}$ represent the training dataset, where $y_i \in \{1,2,\dots,C\}$ and $\mathbf{x}_i \in \mathbb{R}^d$. Let $N_c$ denote the number of samples in each class $c$, and let
\begin{equation}
\boldsymbol{\mu}_c = \frac{1}{N_c} \sum_{i: y_i = c} \mathbf{x}_i
\end{equation}
represent the mean vector of class $c$. The global mean of the dataset is defined as
\begin{equation}
\boldsymbol{\mu} = \frac{1}{N} \sum_{i=1}^{N} \mathbf{x}_i.
\end{equation}

LDA seeks a linear projection that minimizes within-class variance and maximizes between-class separability. To achieve this objective, the within-class scatter matrix $S_W \in \mathbb{R}^{d \times d}$ can be expressed as
\begin{equation}
S_W = \sum_{c=1}^{C} \sum_{i: y_i = c} (\mathbf{x}_i - \boldsymbol{\mu}_c)(\mathbf{x}_i - \boldsymbol{\mu}_c)^\top,
\end{equation}
and the between-class scatter matrix $S_B \in \mathbb{R}^{d \times d}$ is defined as
\begin{equation}
S_B = \sum_{c=1}^{C} N_c (\boldsymbol{\mu}_c - \boldsymbol{\mu})(\boldsymbol{\mu}_c - \boldsymbol{\mu})^\top.
\end{equation}

The generalized eigenvalue problem
\begin{equation}
S_B \mathbf{w} = \lambda S_W \mathbf{w},
\end{equation}
is solved to obtain the optimal projection matrix $W \in \mathbb{R}^{d \times k}$, where $\mathbf{w}$ represents the corresponding eigenvectors and $\lambda$ represents the generalized eigenvalues. The $k$ eigenvectors corresponding to the largest eigenvalues, where $k \leq C - 1$, are selected to construct the projection matrix $W$.

Next, a lower-dimensional, class-discriminative subspace is obtained by projecting each input feature vector $\mathbf{x}_i$ as
\begin{equation}
\mathbf{z}_i = W^\top \mathbf{x}_i, \quad \mathbf{z}_i \in \mathbb{R}^k.
\end{equation}

In the proposed framework, this supervised projection serves two essential purposes. First, by explicitly incorporating label information into the feature transformation, it improves class separability. Second, it produces a compact representation that is compatible with quantum encoding, enabling direct alignment between the number of available qubits and the dimensionality of the projected space. The resulting vectors $\mathbf{z}_i$ are then used as inputs to the quantum feature map for kernel construction.

\subsection{Quantum Kernel Construction}
\begin{figure}
    \centering
    \includegraphics[width=0.99\linewidth]{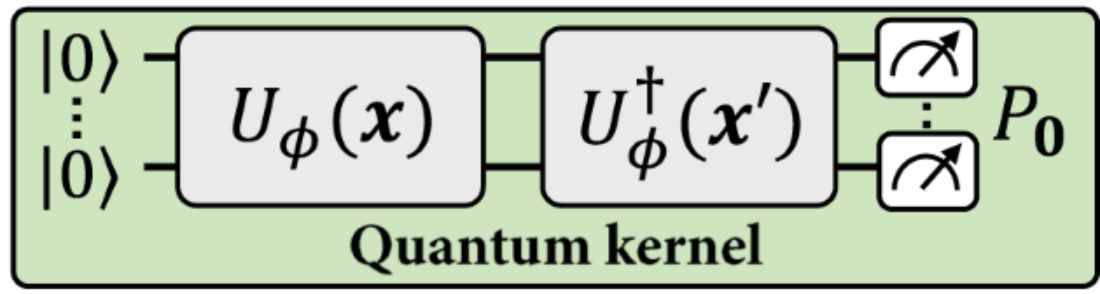}
    \caption{Quantum kernel construction using a parameterized quantum feature map. Input feature vectors are encoded via $U_{\phi}(\cdot)$, and the kernel value is obtained from the measurement probability representing the state overlap.}
    \label{fig:kernel}
\end{figure}

Each malware sample is represented by a low-dimensional vector $\mathbf{z}_i \in \mathbb{R}^k$, where $k \leq C-1$, following supervised feature projection. A parameterized quantum feature map is used to encode these projected features into quantum states, enabling the construction of a quantum kernel that captures nonlinear similarities between malware samples. Fig.~\ref{fig:kernel} illustrates the computation of the proposed quantum kernel.

Let $\mathcal{H}$ represent the Hilbert space of an $n$-qubit quantum system, where the dimensionality of the projected feature space is represented by $n = k$. A unitary transformation
\begin{equation}
U_{\phi}(\mathbf{z}) : \mathbb{R}^k \rightarrow \mathcal{H},
\end{equation}
converts a classical feature vector $\mathbf{z}$ into a quantum state
\begin{equation}
\ket{\phi(\mathbf{z})} = U_{\phi}(\mathbf{z}) \ket{0}^{\otimes n},
\end{equation}
referred to as a quantum feature map.

In this framework, the feature map $U_{\phi}(\mathbf{z})$ is implemented as a parameterized quantum circuit (PQC) composed of entangling operations and single-qubit rotation gates. In particular, rotation gates of the form
\begin{equation}
R_Y(z_j) = \exp\left(-i \frac{z_j}{2} Y\right),
\end{equation}
are applied to the $j$-th qubit to encode each feature component $z_j$. This is followed by a fixed entangling structure repeated for a predetermined number of layers. The expressivity of the quantum feature map is governed by the number of circuit repetitions (reps), which is denoted by $L$.

Given two projected malware samples, $\mathbf{z}_i$ and $\mathbf{z}_j$, the squared fidelity between their respective quantum states defines the quantum kernel as
\begin{equation}
K(\mathbf{z}_i, \mathbf{z}_j) = \left| \braket{\phi(\mathbf{z}_i)}{\phi(\mathbf{z}_j)} \right|^2.
\end{equation}

This fidelity-based kernel function is a valid Mercer kernel, as it is symmetric and positive semi-definite. The kernel value intuitively quantifies the similarity between malware samples in the quantum-induced feature space, where quantum entanglement and interference enable the capture of nonlinear correlations.

Quantum simulation or quantum hardware primitives can be used to estimate the overlap between two quantum states in order to evaluate $K(\mathbf{z}_i, \mathbf{z}_j)$. However, the computational cost of constructing the full kernel matrix
\begin{equation}
\mathbf{K} \in \mathbb{R}^{N \times N}, \quad K_{ij} = K(\mathbf{z}_i, \mathbf{z}_j),
\end{equation}
grows quadratically with the number of training samples $N$. This limitation motivates the scalable kernel approximation method described in the next subsection.

\subsection{Scalable Quantum Kernel Approximation via Nystr\"{o}m Method}

For large-scale datasets, it is computationally prohibitive to directly compute and store the full quantum kernel matrix $\mathbf{K} \in \mathbb{R}^{N \times N}$, as it requires $\mathcal{O}(N^2)$ kernel evaluations and memory. We adopt the Nystr\"{o}m approach to generate a low-rank approximation of the quantum kernel matrix in order to mitigate this limitation and enable learning from all available training samples.

A subset of $M \ll N$ landmark samples selected from the training set is represented by $\mathcal{L} = \{\mathbf{z}_{\ell_1}, \mathbf{z}_{\ell_2}, \dots, \mathbf{z}_{\ell_M}\}$. Using these landmarks, we define the submatrices as
\begin{equation}
\mathbf{K}_{MM} \in \mathbb{R}^{M \times M}, \quad [\mathbf{K}_{MM}]_{ab} = K(\mathbf{z}_{\ell_a}, \mathbf{z}_{\ell_b}),
\end{equation}
and
\begin{equation}
\mathbf{K}_{NM} \in \mathbb{R}^{N \times M}, \quad [\mathbf{K}_{NM}]_{ia} = K(\mathbf{z}_i, \mathbf{z}_{\ell_a}),
\end{equation}
where $K(\cdot,\cdot)$ denotes the fidelity-based quantum kernel defined in the previous subsection.

The Nystr\"{o}m approximation of the full kernel matrix $\mathbf{K}$ is given by
\begin{equation}
\tilde{\mathbf{K}} = \mathbf{K}_{NM} \mathbf{K}_{MM}^{-1} \mathbf{K}_{MN},
\end{equation}
where $\mathbf{K}_{MN} = \mathbf{K}_{NM}^\top$. We apply Tikhonov regularization to $\mathbf{K}_{MM}$ and compute
\begin{equation}
\mathbf{K}_{MM}^{-1} \approx (\mathbf{K}_{MM} + \epsilon \mathbf{I})^{-1},
\end{equation}
where $\epsilon > 0$ is a small regularization constant used to improve numerical stability.

For each sample, the Nystr\"{o}m approximation produces an explicit low-dimensional feature representation. In particular, the Nystr\"{o}m feature mapping $\boldsymbol{\psi}(\mathbf{z}_i) \in \mathbb{R}^M$ is defined as
\begin{equation}
\boldsymbol{\psi}(\mathbf{z}_i) = \mathbf{K}_{MM}^{-1/2} \mathbf{k}_i,
\end{equation}
with
\begin{equation}
\mathbf{k}_i = \left[ K(\mathbf{z}_i, \mathbf{z}_{\ell_1}), \dots, K(\mathbf{z}_i, \mathbf{z}_{\ell_M}) \right]^\top,
\end{equation}
and the approximate kernel can be expressed as the inner product
\begin{equation}
\tilde{K}(\mathbf{z}_i, \mathbf{z}_j) = \boldsymbol{\psi}(\mathbf{z}_i)^\top \boldsymbol{\psi}(\mathbf{z}_j).
\end{equation}

This formulation requires only $\mathcal{O}(NM)$ kernel evaluations and $\mathcal{O}(M^2)$ memory, allowing the quantum kernel approach to scale linearly with the number of training samples $N$. Importantly, the Nystr\"{o}m approximation largely preserves classification performance while maintaining computational tractability by enabling the use of all available training data without the need for subsampling.

\subsection{Multiclass Classification and Training Procedure}
Each malware sample $\mathbf{z}_i$ is mapped to an explicit feature representation $\boldsymbol{\psi}(\mathbf{z}_i) \in \mathbb{R}^{M}$ using the Nystr\"{o}m approximation discussed in the preceding subsection. Let 
\begin{equation}
\mathbf{\Psi} = 
\begin{bmatrix}
\boldsymbol{\psi}(\mathbf{z}_1)^\top \\
\boldsymbol{\psi}(\mathbf{z}_2)^\top \\
\vdots \\
\boldsymbol{\psi}(\mathbf{z}_N)^\top
\end{bmatrix}
\in \mathbb{R}^{N \times M}
\end{equation}
denote the Nystr\"{o}m feature matrix constructed from the training data.

We use a regularized linear ridge classifier in the Nystr\"{o}m-induced feature space to perform multiclass malware family classification. The one-hot encoded label matrix is represented as $\mathbf{Y} \in \mathbb{R}^{N \times C}$, where $Y_{ic} = 1$ if sample $i$ belongs to class $c$ and $Y_{ic} = 0$ otherwise. Let $\mathbf{W} \in \mathbb{R}^{M \times C}$ represent the weight matrix, $||\cdot||_F$ denote the Frobenius norm, and $\lambda > 0$ be a regularization parameter that controls model complexity. The classifier is trained by minimizing the ridge regression objective defined as
\begin{equation}
\mathcal{L}(\mathbf{W}) = ||\mathbf{\Psi}\mathbf{W} - \mathbf{Y}||_F^2 + \lambda ||\mathbf{W}||_F^2.
\end{equation}

A closed-form solution to the optimization problem is given by
\begin{equation}
\mathbf{W}^{\ast} = \left( \mathbf{\Psi}^\top \mathbf{\Psi} + \lambda \mathbf{I} \right)^{-1} \mathbf{\Psi}^\top \mathbf{Y},
\end{equation}
where $\mathbf{I}$ represents the identity matrix of the appropriate dimension. Because the dimensionality of the optimization problem depends on the number of landmarks $M$ rather than the number of samples $N$, this formulation enables efficient training even for large-scale datasets.

During inference, a test sample $\mathbf{z}_{\mathrm{test}}$ is first mapped to its Nystr\"{o}m feature representation $\boldsymbol{\psi}(\mathbf{z}_{\mathrm{test}})$ using the same LDA transformation learned from the training data. The predicted class label is then obtained as
\begin{equation}
\hat{y} = \arg\max_{c \in \{1,\dots,C\}} \left( \boldsymbol{\psi}(\mathbf{z}_{\mathrm{test}})^\top \mathbf{W}^{\ast} \right)_c.
\end{equation}

Thus, the entire training pipeline consists of feature extraction, supervised projection, Nystr\"{o}m feature construction, quantum kernel evaluation on landmark samples, and closed-form ridge regression. For large-scale multiclass malware family classification, this approach enables the proposed framework to leverage expressive quantum kernels while maintaining scalability and computational efficiency.

\begin{algorithm}[t]
\footnotesize
\caption{Scalable Quantum Kernel Framework for Malware Family Classification}
\label{alg:qsml_malware}
\begin{algorithmic}[1]
\REQUIRE Training set $\mathcal{D}_{\mathrm{train}}=\{(\mathbf{x}_i,y_i)\}_{i=1}^{N}$, test set $\mathcal{D}_{\mathrm{test}}=\{(\mathbf{x}_j,y_j)\}_{j=1}^{N_{\mathrm{test}}}$; number of classes $C$; qubits $n$; circuit repetitions $L$; landmark size $M$; regularization $\lambda$; Nystr\"{o}m regularizer $\epsilon$
\ENSURE Predicted labels $\{\hat{y}_j\}_{j=1}^{N_{\mathrm{test}}}$ and trained model parameters $(W_{\mathrm{LDA}},\mathcal{L},\mathbf{K}_{MM}^{-1/2},\mathbf{W}^{\ast})$

\STATE Extract raw PE-based features for all samples and form matrices $\mathbf{X}_{\mathrm{train}}\in\mathbb{R}^{N\times d}$, $\mathbf{X}_{\mathrm{test}}\in\mathbb{R}^{N_{\mathrm{test}}\times d}$ \label{line:feat}
\STATE Standardize features using training statistics: $\tilde{\mathbf{X}}_{\mathrm{train}},\tilde{\mathbf{X}}_{\mathrm{test}}$ \label{line:scale}

\STATE Compute LDA projection matrix $W_{\mathrm{LDA}}\in\mathbb{R}^{d\times n}$ using $\tilde{\mathbf{X}}_{\mathrm{train}}$ and labels $\mathbf{y}_{\mathrm{train}}$ \label{line:lda_fit}
\STATE Project data: $\mathbf{Z}_{\mathrm{train}}=\tilde{\mathbf{X}}_{\mathrm{train}}W_{\mathrm{LDA}}$, $\mathbf{Z}_{\mathrm{test}}=\tilde{\mathbf{X}}_{\mathrm{test}}W_{\mathrm{LDA}}$ \label{line:lda_apply}

\STATE Select landmark indices $\{\ell_1,\dots,\ell_M\}$ from training data; define $\mathcal{L}=\{\mathbf{z}_{\ell_m}\}_{m=1}^{M}$ \label{line:landmarks}

\STATE Initialize $\mathbf{K}_{MM}\in\mathbb{R}^{M\times M}$ \label{line:init_kmm}
\FOR{$a=1$ \TO $M$} \label{line:for_kmm_a}
    \FOR{$b=1$ \TO $M$} \label{line:for_kmm_b}
        \STATE Prepare $\ket{\phi(\mathbf{z}_{\ell_a})}$ and $\ket{\phi(\mathbf{z}_{\ell_b})}$ using $n$ qubits and $L$ repetitions \label{line:pqc_kmm}
        \STATE Compute fidelity kernel $[\mathbf{K}_{MM}]_{ab}=|\langle\phi(\mathbf{z}_{\ell_a})|\phi(\mathbf{z}_{\ell_b})\rangle|^2$ \label{line:fid_kmm}
    \ENDFOR \label{line:endfor_kmm_b}
\ENDFOR \label{line:endfor_kmm_a}

\STATE Regularize and factorize: $\mathbf{K}_{MM}\leftarrow \mathbf{K}_{MM}+\epsilon\mathbf{I}$, compute $\mathbf{K}_{MM}^{-1/2}$ \label{line:kmm_invhalf}

\STATE Initialize Nystr\"{o}m feature matrix $\mathbf{\Psi}\in\mathbb{R}^{N\times M}$ \label{line:init_psi}
\FOR{$i=1$ \TO $N$} \label{line:for_knm_i}
    \STATE Compute $\mathbf{k}_i=[K(\mathbf{z}_i,\mathbf{z}_{\ell_1}),\dots,K(\mathbf{z}_i,\mathbf{z}_{\ell_M})]^\top$ via quantum kernel evaluations \label{line:knm_row}
    \STATE Set $\mathbf{\Psi}_{i:}^\top \leftarrow \mathbf{K}_{MM}^{-1/2}\mathbf{k}_i$ \label{line:psi_row}
\ENDFOR \label{line:endfor_knm_i}

\STATE Build one-hot labels $\mathbf{Y}\in\mathbb{R}^{N\times C}$ from $\mathbf{y}_{\mathrm{train}}$ \label{line:onehot}
\STATE Train multiclass ridge: $\mathbf{W}^{\ast}=(\mathbf{\Psi}^\top\mathbf{\Psi}+\lambda\mathbf{I})^{-1}\mathbf{\Psi}^\top\mathbf{Y}$ \label{line:ridge}

\STATE Initialize predictions $\{\hat{y}_j\}_{j=1}^{N_{\mathrm{test}}}$ \label{line:init_pred}
\FOR{$j=1$ \TO $N_{\mathrm{test}}$} \label{line:for_test}
    \STATE Compute $\mathbf{k}^{(\mathrm{test})}_j=[K(\mathbf{z}^{(\mathrm{test})}_j,\mathbf{z}_{\ell_1}),\dots,K(\mathbf{z}^{(\mathrm{test})}_j,\mathbf{z}_{\ell_M})]^\top$ \label{line:test_knm}
    \STATE Compute Nystr\"{o}m features $\boldsymbol{\psi}(\mathbf{z}^{(\mathrm{test})}_j)=\mathbf{K}_{MM}^{-1/2}\mathbf{k}^{(\mathrm{test})}_j$ \label{line:test_psi}
    \STATE Predict $\hat{y}_j=\arg\max_{c\in\{1,\dots,C\}} \left(\boldsymbol{\psi}(\mathbf{z}^{(\mathrm{test})}_j)^\top\mathbf{W}^{\ast}\right)_c$ \label{line:predict}
\ENDFOR \label{line:endfor_test}

\RETURN $\{\hat{y}_j\}_{j=1}^{N_{\mathrm{test}}}$ and $(W_{\mathrm{LDA}},\mathcal{L},\mathbf{K}_{MM}^{-1/2},\mathbf{W}^{\ast})$ \label{line:return}
\end{algorithmic}
\end{algorithm}

\subsection{Algorithm Description}
Algorithm~\ref{alg:qsml_malware} describes the end-to-end pipeline for scalable malware family classification utilizing a quantum fidelity kernel and Nystr\"{o}m approximation. To create the training and test feature matrices \eqref{line:feat}, static PE-based features are first extracted from each malware executable. All features are normalized using statistics computed from the training set, and the same transformation is applied to the test set \eqref{line:scale}, as malware characteristics may vary significantly in scale (e.g., file size versus entropy statistics).

The standardized training data \eqref{line:lda_fit} is used to construct a supervised Linear Discriminant Analysis (LDA) projection matrix in order to match the limited number of available qubits and reduce dimensionality in a label-aware manner. After that, the training and test features are projected onto an $n$-dimensional space, yielding $\mathbf{Z}_{\mathrm{train}}$ and $\mathbf{Z}_{\mathrm{test}}$ \eqref{line:lda_apply}. The quantum kernel is then approximated at scale using the landmark set $\mathcal{L}$ \eqref{line:landmarks}, which is created by selecting a subset of $M$ representative training samples as landmarks.

A circuit with $n$ qubits and $L$ repetitions implements a parameterized quantum feature map, which is used to construct the quantum kernel. By calculating the squared state fidelity between each pair of landmark samples \eqref{line:pqc_kmm}--\eqref{line:fid_kmm}, the method first computes the landmark-to-landmark kernel matrix $\mathbf{K}_{MM}$ \eqref{line:init_kmm}. Since $\mathbf{K}_{MM}$ must be numerically stable and invertible, it is factorized to obtain $\mathbf{K}_{MM}^{-1/2}$ \eqref{line:kmm_invhalf} after being regularized with a small diagonal term. This factor enables an explicit Nystr\"{o}m feature representation.

For each sample, we compute the kernel similarities between the sample and the $M$ landmarks, yielding a kernel vector $\mathbf{k}_i$ \eqref{line:knm_row}, which is subsequently converted into an $M$-dimensional Nystr\"{o}m feature vector via $\mathbf{K}_{MM}^{-1/2}$ \eqref{line:psi_row}. The Nystr\"{o}m feature matrix $\mathbf{\Psi}$ \eqref{line:init_psi}, which approximates the quantum kernel in low-rank form, is obtained by collecting these rows. A multiclass ridge regression model is trained in closed form \eqref{line:ridge} using the one-hot label matrix \eqref{line:onehot}. This model is computationally efficient because its complexity mainly depends on $M$ rather than the total dataset size $N$.

Each test sample is similarly compared to the landmark set during inference to create a kernel vector \eqref{line:test_knm}, which is then transformed into Nystr\"{o}m features \eqref{line:test_psi} and classified by selecting the class with the highest ridge score \eqref{line:predict}. Finally, the method outputs the classifier parameters \eqref{line:return}, the learned projection, landmark set, Nystr\"{o}m factors, and predicted labels. This framework enables scalable learning from all available training data while maintaining the expressive capacity of the quantum fidelity kernel.

\noindent \textbf{Computational Complexity Analysis.} We examine the computational complexity of the proposed approach with respect to the number of qubits $n$, feature dimension $d$, Nystr\"{o}m landmarks $M$, number of training samples $N$, and circuit repetitions $L$. In the worst case, feature preprocessing and supervised projection using LDA require $O(N d^2)$ time and are performed once during training. The dominant cost arises from quantum kernel construction: $O(M^2 \cdot C_{\text{q}})$ is required to compute the landmark kernel matrix. Evaluating kernel similarities between all training samples and landmarks requires $O(NM \cdot C_{\text{q}})$ operations, where the cost of simulating a parameterized quantum circuit with $n$ qubits and $L$ repetitions is denoted by $C_{\text{q}}$.

Nystr\"{o}m feature construction involves the eigen-decomposition of the $M \times M$ landmark kernel matrix, which in the worst case requires $O(M^3)$ operations. This is followed by feature projection, which requires $O(NM^2)$ operations. Finally, a linear system with complexity $O(M^3 + NM^2)$ must be solved to train the multiclass ridge classifier on the Nystr\"{o}m features. Crucially, the proposed strategy reduces the dominant kernel-related complexity to $O(NM)$ by replacing the full $O(N^2)$ kernel computation with the Nystr\"{o}m approximation. This enables scalable learning on large malware datasets while preserving expressive quantum similarity modeling.

\begin{figure*} [h]
    \centering
    \includegraphics[width=0.99\linewidth]{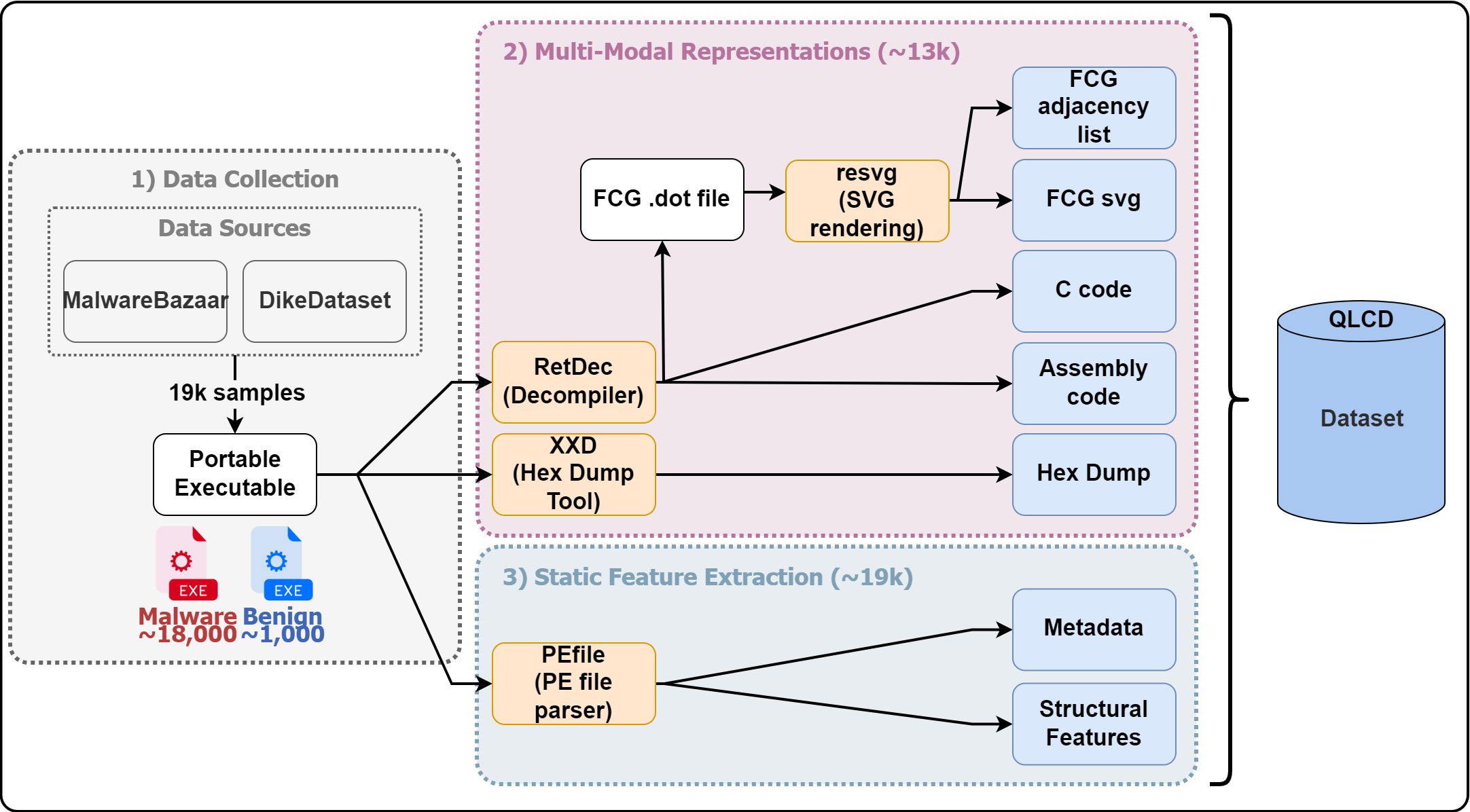}
    \caption{QLCD dataset construction workflow}
    \label{fig:qlc}
\end{figure*}

\section{Experiments}
This section presents the experimental results for the proposed scalable QKML framework for malware family classification. The experiments are designed to evaluate classification performance, scalability, and robustness, while providing a fair comparison with representative classical machine learning baselines.

\subsection{Dataset Construction and Description}

To construct QLCD (Quantum Learning Code Dataset), we aggregated Portable Executable (PE) files from two primary sources: the DikeDataset \cite{DikeDataset2023} and MalwareBazaar \cite{MalwareBazaar}. The DikeDataset contributed 982 benign and 8,970 malicious samples, forming the initial corpus. To ensure coverage of recent and evolving threats, we further collected 10,000 additional PE files from MalwareBazaar spanning the 2025–2026 period. This expansion allows the dataset to capture contemporary malware campaigns and newly emerging variants. A subset of malicious binaries could not be fully processed due to adversarial characteristics such as advanced obfuscation, binary packing, encryption, and structural corruption designed to hinder static analysis tools.

Moreover, the dataset generation pipeline follows two parallel branches: multi-modal representation extraction and static feature extraction, as illustrated in Fig.~\ref{fig:qlc}. For multi-modal representation generation, we employed the RetDec decompiler to derive higher-level program abstractions from raw binaries. Specifically, RetDec was used to produce decompiled C source code, assembly instructions, and Function Call Graphs (FCGs) in DOT format. These graph representations were further processed into adjacency structures and visual formats. In parallel, the XXD utility was used to generate raw hexadecimal dumps to preserve byte-level representations. Due to the complexity and obfuscation of certain malware samples, this stage experienced the highest attrition, resulting in approximately 13,000 samples with complete multi-view representations.

In the second branch, we performed static structural analysis using the Python-based \texttt{pefile} library. This process enabled robust parsing of PE headers and internal structures to extract metadata and structural artifacts, including imported APIs, extracted strings, section headers, entropy measurements, and other header-level attributes. Static parsing proved more resilient to obfuscation compared to decompilation, allowing approximately 19,000 samples to retain usable static feature representations. These artifacts support structural analysis and metadata-driven reasoning tasks.

Finally, malicious samples were annotated using available ground-truth labels and validated heuristics to assign malware family names and broader behavioral categories. While a limited subset contains partial annotations, the majority of samples retain complete labeling information. The resulting QLCD dataset therefore aligns raw binaries, multi-view program representations, structural features, and family-level annotations within a unified framework for early static malware family attribution.

\subsubsection{Dataset Processing}

A large malware dataset composed of Portable Executable (PE) files collected from real-world threat intelligence sources is used for the experiments. The \texttt{malware\_family} attribute, which is obtained from carefully curated cyber threat intelligence metadata, is used to label each sample. Static structural features are extracted directly from PE headers and section information without requiring dynamic execution, resulting in efficient and scalable feature extraction suitable for large-scale deployments.

Malware families with fewer than 50 samples are excluded to ensure reliable multiclass learning. Following filtering, the final dataset contains a total of 23 classes, comprising 18{,}836 samples from 22 malware families and one benign class. In addition to benign software samples, the malware families include \textit{AgentTesla, AsyncRAT, CobaltStrike, CoinMiner, Conti, DarkComet, Emotet, GuLoader, LockBit, LokiBot, TrickBot, XMRig, locker, mediyes, unknown, winwebsec, zbot, and zeroaccess}, as shown in Table~\ref{tab:dataset_split}. The dataset exhibits significant class imbalance, with sample counts ranging from 4,400 (winwebsec) to 74 (LokiBot). This variation reflects realistic operational cybersecurity environments in which certain malware families are far more prevalent than others. Furthermore, due to code reuse, obfuscation, and packing techniques, many malware families share structural characteristics, which makes precise family-level classification particularly challenging.

\begin{table}[t]
\centering
\footnotesize
\caption{Dataset composition showing the number of training and testing samples for each class after preprocessing and stratified splitting (25\% per class with rounding). The total dataset contains 18,836 samples across 23 classes (22 malware families and one benign class).}
\label{tab:dataset_split}
\setlength{\tabcolsep}{4pt}
\begin{tabular}{lccc}
\toprule
\textbf{Class Label} & \textbf{Training Samples} & \textbf{Testing Samples} & \textbf{Total Samples} \\
\midrule
benign & 731 & 244 & 975 \\
\midrule
AgentTesla & 80 & 26 & 106 \\
AsyncRAT & 668 & 222 & 890 \\
CobaltStrike & 183 & 61 & 244 \\
CoinMiner & 226 & 75 & 301 \\
Conti & 713 & 238 & 951 \\
DarkComet & 134 & 44 & 178 \\
Emotet & 138 & 46 & 184 \\
GuLoader & 584 & 195 & 779 \\
LockBit & 76 & 26 & 102 \\
LokiBot & 56 & 18 & 74 \\
Metasploit & 193 & 64 & 257 \\
NjRAT & 289 & 96 & 385 \\
Petya & 64 & 21 & 85 \\
Remcos & 86 & 29 & 115 \\
TrickBot & 61 & 20 & 81 \\
XMRig & 213 & 71 & 284 \\
Locker & 232 & 78 & 310 \\
Mediyes & 1088 & 362 & 1450 \\
Unknown & 2921 & 974 & 3895 \\
Winwebsec & 3300 & 1100 & 4400 \\
Zbot & 1575 & 525 & 2100 \\
Zeroaccess & 518 & 172 & 690 \\
\midrule
\textbf{Total} & \textbf{14,129} & \textbf{4,707} & \textbf{18,836} \\
\bottomrule
\end{tabular}
\end{table}


\subsection{Experimental Setup}
The experimental setup used to evaluate the proposed scalable QKML framework is described in this subsection. The dataset, evaluation metrics, baseline methods, implementation settings, and computing environment are described in detail. To ensure fairness and reproducibility, the same data splits and feature representations are used in all experiments.

\subsubsection{Baseline Methods}
We evaluate the performance of the proposed quantum learning framework by comparing it with commonly used classical machine learning baselines that are widely employed in malware classification. To ensure a fair comparison, all baseline models use the same training and test splits as the proposed method and operate on the same feature representation.

\begin{figure}
    \centering
    \includegraphics[width=0.99\linewidth]{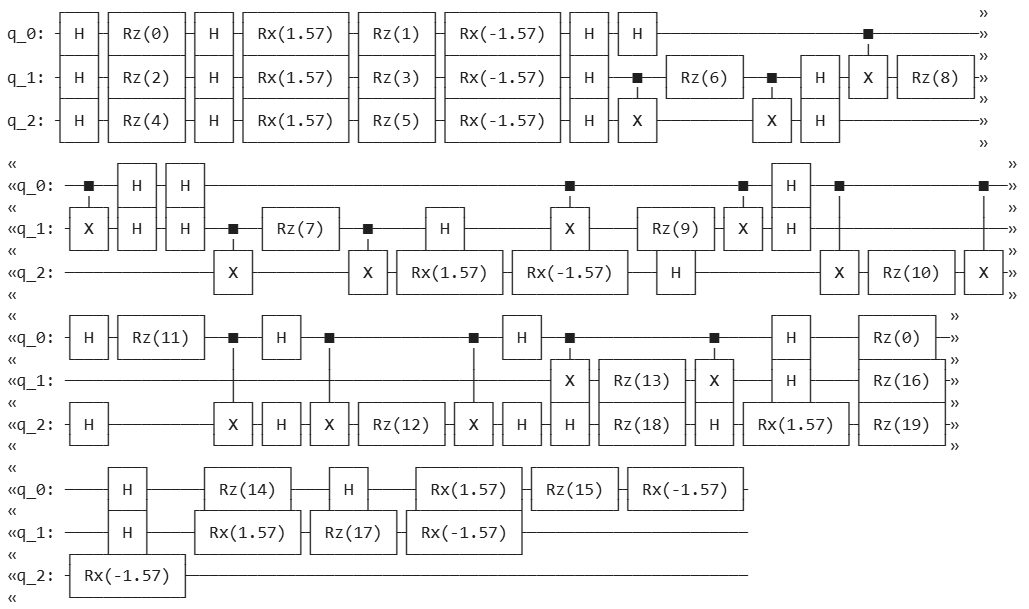}
    \caption{A three-qubit quantum feature-map circuit for kernel evaluation, consisting of layered single-qubit rotations (RZ/RX), Hadamard gates, entangling CX operations, and a final measurement.}
    \label{fig:circuit}
\end{figure}

The selected baselines include traditional kernel-based techniques and linear ridge regression. Specifically, we examine ridge regression combined with the classical Nystr\"{o}m kernel approximation, which provides a scalable alternative to full kernel learning. Under identical experimental conditions, these baselines provide strong benchmarks to assess whether quantum kernel learning offers measurable improvements over classical methods. The circuit used for the quantum feature map is shown in Fig.~\ref{fig:circuit}. 

\subsubsection{Implementation Details and Hyperparameters}
A hybrid classical–quantum pipeline is used to implement the proposed framework. Feature preprocessing includes one-hot encoding of categorical attributes, z-score normalization, and median imputation for missing numerical values. Using Linear Discriminant Analysis (LDA), supervised dimensionality reduction is performed by projecting the initial feature space to a dimension equivalent to the number of qubits.

Quantum feature encoding is implemented using a parameterized quantum circuit composed of ring-based entangling operations and single-qubit rotation gates. To improve expressivity, the circuit is repeated $L=4$ times with the number of qubits set to $n=8$. Before being mapped to the interval $[-\pi, \pi]$ for angle encoding, projected features are robustly scaled and clipped.

Using $M=8000$ landmark samples selected from the entire training set, the Nystr\"{o}m approximation is employed to enable scalable kernel learning. Statevector simulation is used to compute a fidelity-based quantum kernel, and multiclass ridge regression is trained on the resulting Nystr\"{o}m features using all available training samples. The ridge regularization parameter is set to $\lambda = 5 \times 10^{-5}$. Unless otherwise specified, all hyperparameters remain fixed throughout the experiments.

\subsubsection{Computational Environment}
All experiments are conducted on a standard computing platform using a Python-based implementation. A statevector-based quantum circuit simulator is used to evaluate quantum kernels, while CPU resources are used for classical preprocessing and optimization. This configuration ensures complete reproducibility without requiring specialized quantum hardware and reflects realistic near-term conditions for hybrid QML.

\subsection{Simulation Results: Ablation Study}
\begin{figure}[t]
    \centering
    \includegraphics[width=\linewidth]{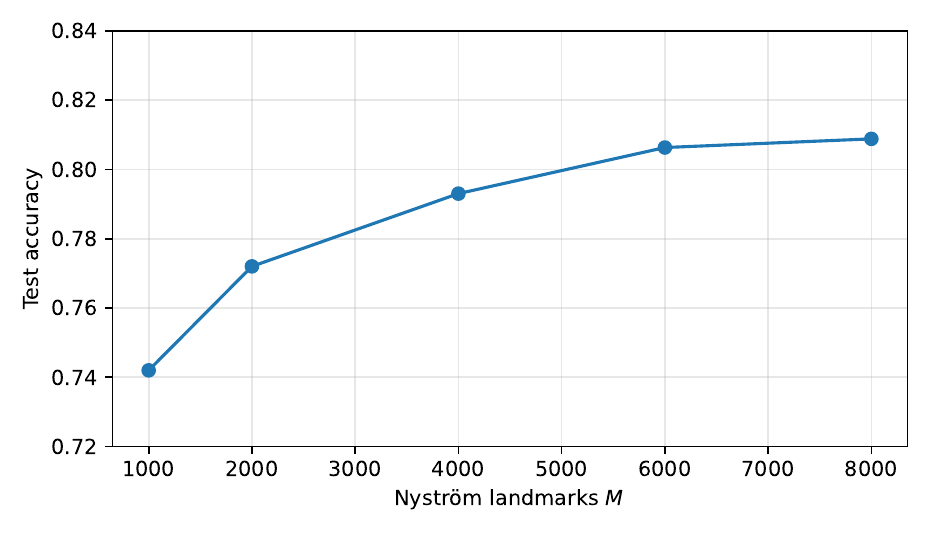}
    \caption{Test accuracy as a function of the number of Nystr\"{o}m landmarks $M$ for the proposed fidelity-kernel quantum learning framework.}
    \label{fig:acc_vs_landmarks}
\end{figure}

\textbf{Scalability Analysis.} The scalability behavior of the proposed quantum learning framework as the number of Nystr\"{o}m landmark samples $M$ increases is shown in Fig.~\ref{fig:acc_vs_landmarks}. The model achieves an accuracy of roughly 74\% when using a small number of landmarks ($M=1000$), indicating that the coarse kernel approximation captures only a limited amount of structural information. The accuracy gradually improves to about 77\% and 80\% as $M$ increases to 2000 and 4000, respectively, showing that a richer low-rank approximation produces more class-discriminative kernel features. The accuracy reaches approximately 81.8\% at $M=8000$ and begins to saturate, indicating diminishing returns beyond this point. This trend demonstrates a clear trade-off between predictive performance and computational efficiency, and it confirms that the Nystr\"{o}m approximation enables scalable quantum kernel learning while maintaining high accuracy on large-scale malware family classification tasks.

\begin{table}[t]
\centering
\caption{Ablation study on projection strategy before quantum encoding (fixed $n=8$, $L=4$, $M=8000$, test=4707).}
\label{tab:ablation_projection}
\begin{tabular}{lcc}
\toprule
\textbf{Projection} & \textbf{Accuracy} & \textbf{Loss} \\
\midrule
SVD (unsupervised) & 0.7883 & 2.5277 \\
LDA (supervised, ours) & 0.8088 & 2.4841 \\
\bottomrule
\end{tabular}
\end{table}

\textbf{Projection Ablation (LDA vs.\ SVD).} The effect of the projection step used prior to quantum feature encoding is evaluated in Table~\ref{tab:ablation_projection}. With the same quantum and Nystr\"{o}m parameters ($n=8$ qubits, $L=4$ circuit repetitions, and $M=8000$ landmarks), the accuracy decreases from 80.88\% to 78.83\% when supervised LDA is replaced with an unsupervised SVD projection. This difference indicates that label-aware dimensionality reduction prior to quantum encoding improves class separability, resulting in more discriminative quantum kernel similarities. Furthermore, LDA achieves a slightly lower cross-entropy loss (2.4841 vs.\ 2.5277), indicating both higher classification accuracy and improved confidence calibration.

\begin{figure}[t]
    \centering
    \includegraphics[width=\linewidth]{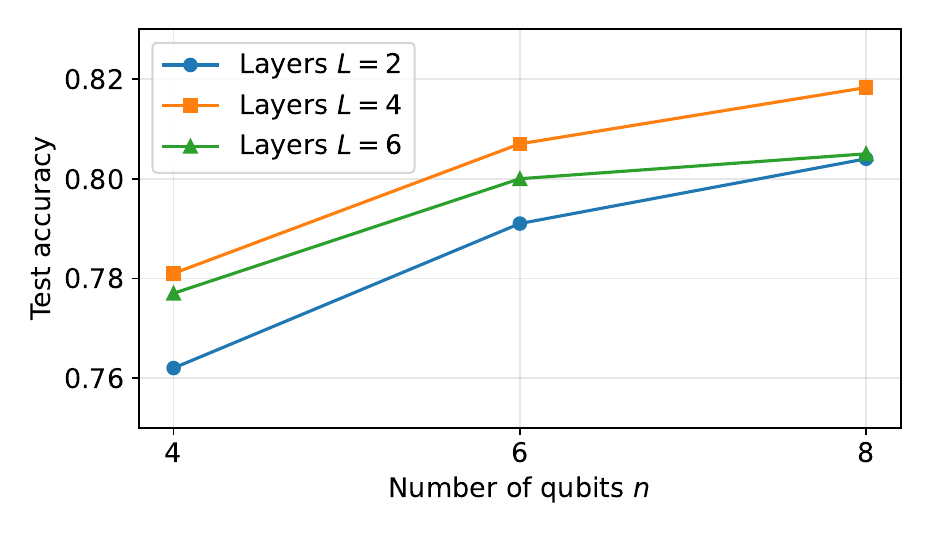}
    \caption{Test accuracy as a function of the number of qubits $n$ and circuit repetitions $L$ for the proposed quantum kernel framework.}
    \label{fig:qubits_reps}
\end{figure}
\begin{figure*}[t]
    \centering
    \begin{subfigure}[t]{0.19\textwidth}
        \centering
        \includegraphics[width=\linewidth, trim=50 50 50 20, clip]{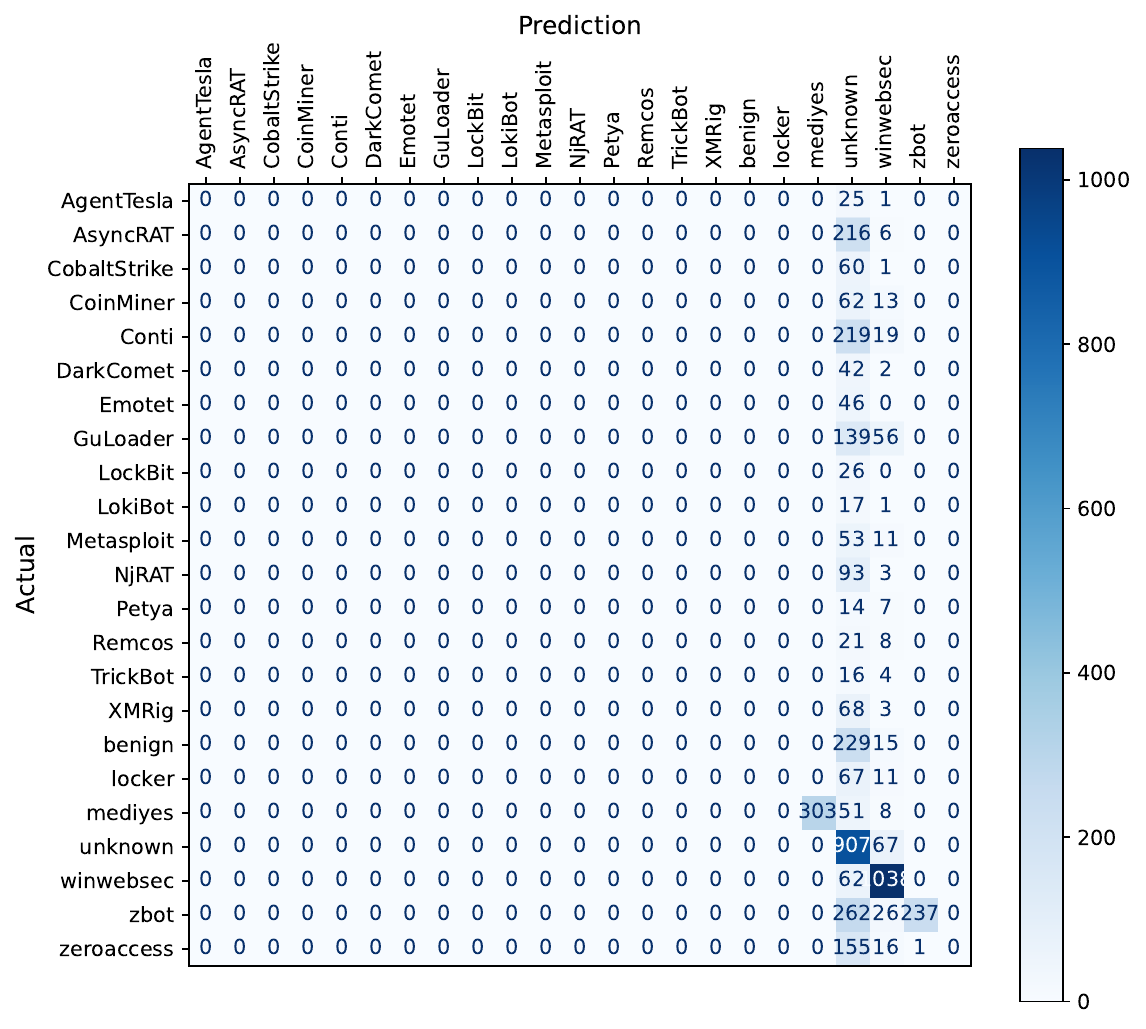}
        \caption{Softmax Regression}
        \label{fig:cm_softmax}
    \end{subfigure}
    \hfill
    \begin{subfigure}[t]{0.19\textwidth}
        \centering
        \includegraphics[width=\linewidth, trim=50 50 50 20, clip]{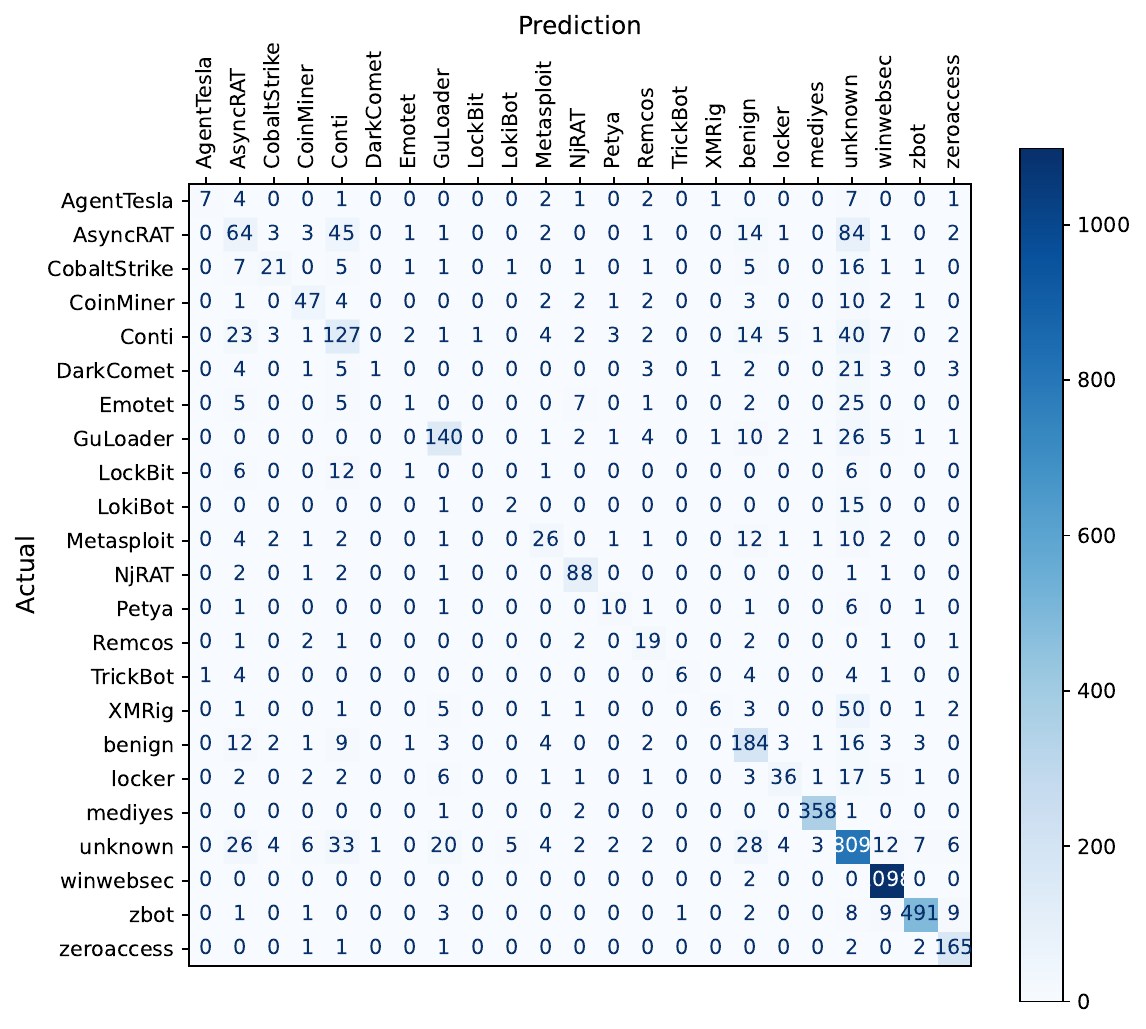}
        \caption{Linear SVM}
        \label{fig:cm_svm}
    \end{subfigure}
    \hfill
    \begin{subfigure}[t]{0.19\textwidth}
        \centering
        \includegraphics[width=\linewidth, trim=50 50 50 20, clip]{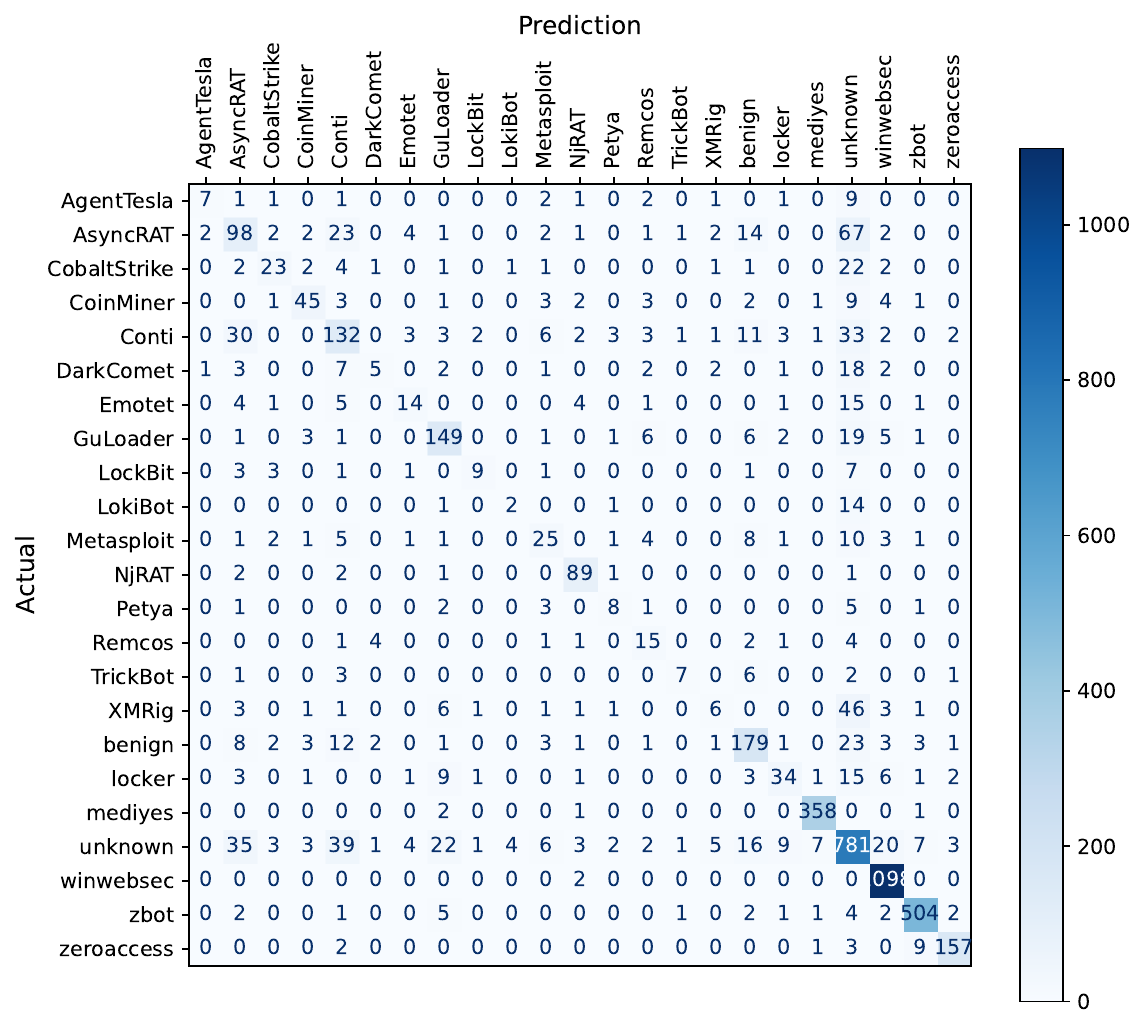}
        \caption{KNN ($k=15$)}
        \label{fig:cm_knn}
    \end{subfigure}
    \hfill
    \begin{subfigure}[t]{0.19\textwidth}
        \centering
        \includegraphics[width=\linewidth, trim=50 50 50 20, clip]{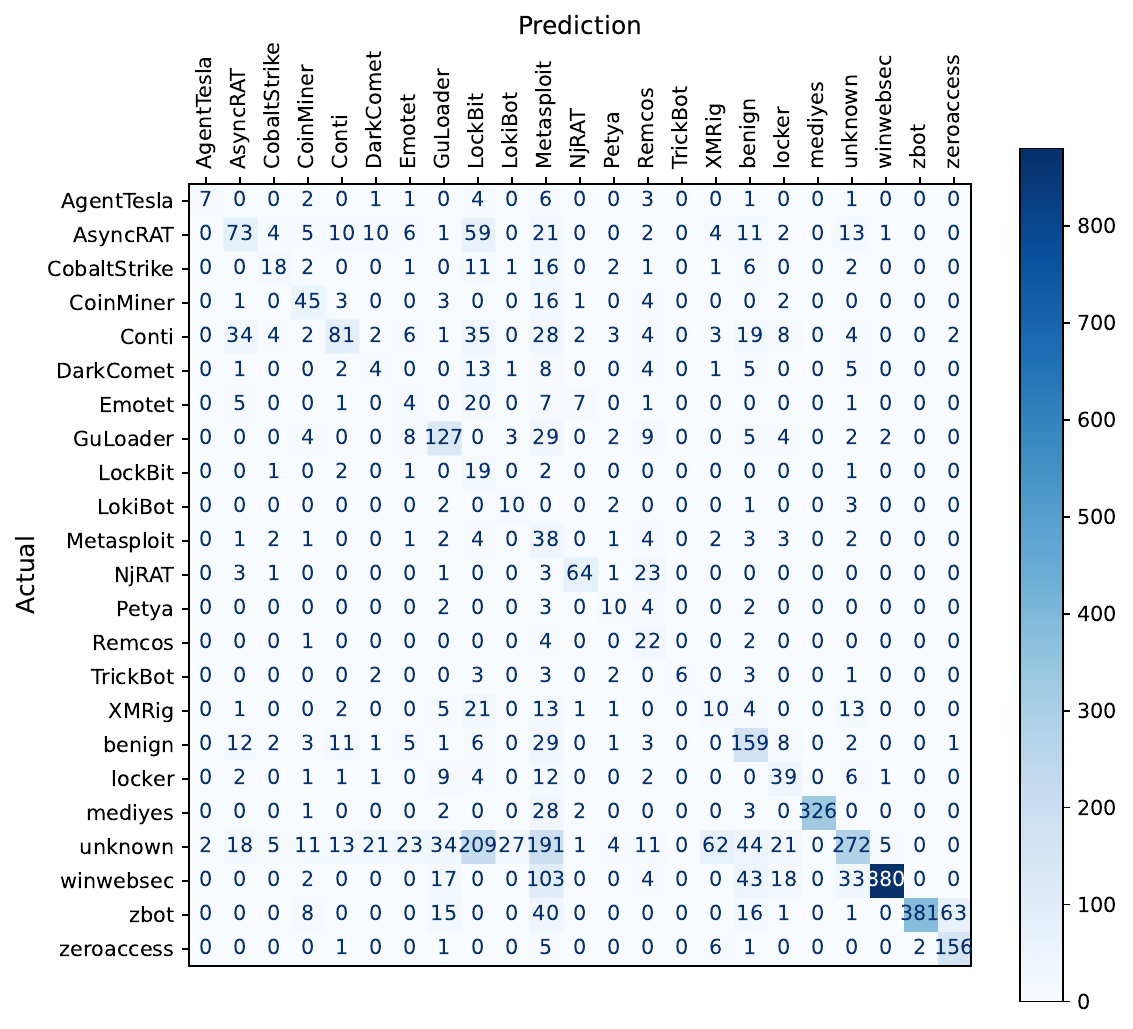}
        \caption{Gaussian NB}
        \label{fig:cm_gnb}
    \end{subfigure}
    \hfill
    \begin{subfigure}[t]{0.19\textwidth}
        \centering
        \includegraphics[width=\linewidth, trim=50 50 50 20, clip]{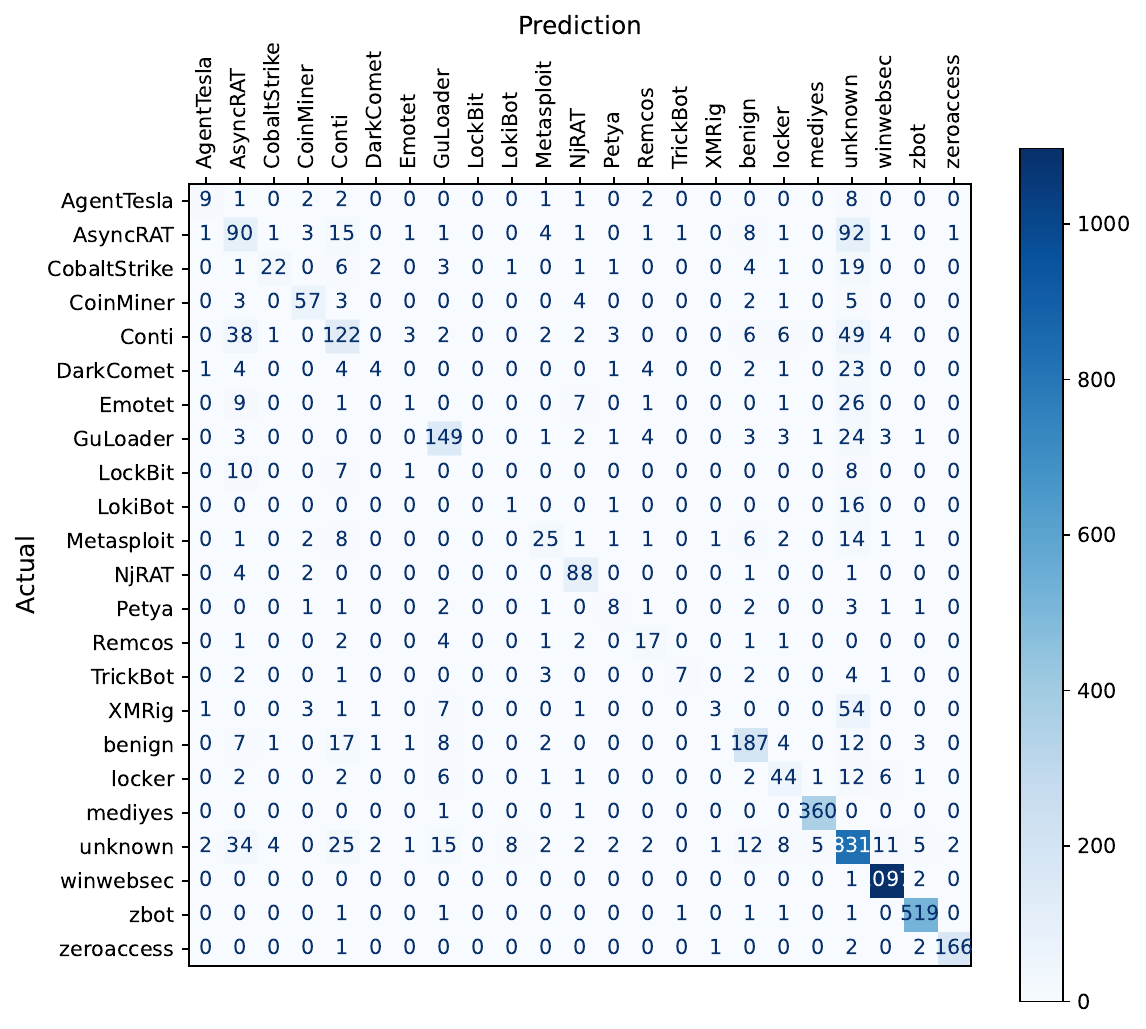}
        \caption{Quantum Kernel}
        \label{fig:cm_quantum}
    \end{subfigure}

    \caption{
    Confusion matrices for multi-class malware family classification using classical and quantum models.
    Classical baselines include Softmax Regression, Linear SVM, KNN, and Gaussian Naïve Bayes,
    while the proposed quantum approach employs a fidelity-based quantum kernel with Nystr\"{o}m approximation.
    All matrices are visualized using color intensity only (without numeric annotations) for clarity,
    scaled to a common evaluation size of 18,836 samples.
    }
    \label{fig:confusion_matrices_all}
\end{figure*}

\textbf{Sensitivity to Qubits and Circuit Layers.} Fig.~\ref{fig:qubits_reps} shows the impact of quantum circuit expressivity on classification performance by varying the number of qubits $n$ and circuit repetitions $L$. Accuracy consistently increases with the number of qubits for both settings, suggesting that class-discriminative structure is better captured by higher-dimensional quantum feature embeddings. All qubit configurations perform better when the circuit depth is increased from $L=2$ to $L=4$. At $n=8$ and $L=4$, the highest accuracy of approximately 81.8\% is achieved. A practical trade-off between circuit complexity and predictive accuracy in near-term quantum kernel learning is observed, with the performance gains gradually saturating as expressivity increases.

\begin{table}[t]
\centering
\caption{Effect of ridge regularization parameter $\lambda$ on classification performance (fixed $n=8$, $L=4$, $M=8000$).}
\label{tab:ablation_lambda}
\begin{tabular}{ccc}
\toprule
\textbf{$\lambda$} & \textbf{Accuracy} & \textbf{Loss} \\
\midrule
$1\times10^{-3}$ & 0.8046 & 2.6124 \\
$1\times10^{-4}$ & 0.8017 & 2.5310 \\
$5\times10^{-5}$ & 0.8088 & 2.4841 \\
\bottomrule
\end{tabular}
\end{table}

\textbf{Regularization Robustness.} Table~\ref{tab:ablation_lambda} examines how sensitive the proposed framework is to the ridge regularization parameter $\lambda$ used in multiclass classification. The cross-entropy loss decreases and the classification accuracy increases from 80.46\% to 80.88\% as $\lambda$ decreases from $10^{-3}$ to $5\times10^{-5}$. This pattern indicates that moderate regularization effectively balances model capacity and numerical stability when learning from Nystr\"{o}m-approximated quantum features. Crucially, performance remains stable across a range of $\lambda$ values, indicating that the proposed approach is robust to regularization choices.

\subsection{Simulation Results: Per-class Performance Analysis} 
To account for the multiclass and imbalanced nature of the malware dataset, we also examine class-balanced performance in addition to overall accuracy. The proposed framework obtains a macro F1-score of 0.53, a macro recall of 0.54, and a macro precision of approximately 0.55 using macro-averaged metrics. These results indicate that the model exhibits balanced predictive performance across malware families rather than being dominated by a few frequent classes.

The confusion matrix further shows that most misclassifications occur between malware families that share structural similarities, such as loaders, packers, or obfuscation techniques. Families with distinctive entropy patterns and PE features are generally classified with high accuracy, whereas polymorphic or highly obfuscated families are more likely to be confused. This behavior is consistent with real-world malware analysis scenarios, where semantic similarity across families presents a fundamental challenge. Despite class imbalance and high inter-family similarity, the macro-level results indicate that the proposed quantum kernel framework generalizes well across diverse malware families.

\begin{table}[t]
\centering
\small
\caption{Per-family classification performance of the proposed quantum learning framework on the malware family classification task. Precision, recall, and F1-score are reported for each malware family using the stratified test set (25\% split, total test samples = 4707).}
\label{tab:per_class_results}
\begin{tabular}{lcccc}
\toprule
\textbf{Malware Family} & \textbf{Precision} & \textbf{Recall} & \textbf{F1-score} & \textbf{Support} \\
\midrule
benign       & 0.782 & 0.766 & 0.774 & 244 \\
\midrule
AgentTesla   & 0.643 & 0.346 & 0.450 & 26 \\
AsyncRAT     & 0.429 & 0.405 & 0.417 & 222 \\
CobaltStrike & 0.759 & 0.361 & 0.489 & 61 \\
CoinMiner    & 0.814 & 0.760 & 0.786 & 75 \\
Conti        & 0.557 & 0.513 & 0.534 & 238 \\
DarkComet    & 0.400 & 0.091 & 0.148 & 44 \\
Emotet       & 0.125 & 0.022 & 0.037 & 46 \\
GuLoader     & 0.749 & 0.764 & 0.756 & 195 \\
LockBit      & 0.000 & 0.000 & 0.000 & 26 \\
LokiBot      & 0.100 & 0.056 & 0.071 & 18 \\
Metasploit   & 0.581 & 0.391 & 0.467 & 64 \\
NjRAT        & 0.772 & 0.917 & 0.838 & 96 \\
Petya        & 0.444 & 0.381 & 0.410 & 21 \\
Remcos       & 0.515 & 0.586 & 0.548 & 29 \\
TrickBot     & 0.778 & 0.350 & 0.483 & 20 \\
XMRig        & 0.429 & 0.042 & 0.077 & 71 \\
Locker       & 0.595 & 0.564 & 0.579 & 78 \\
Mediyes      & 0.981 & 0.994 & 0.988 & 362 \\
Unknown      & 0.690 & 0.853 & 0.763 & 974 \\
Winwebsec    & 0.975 & 0.997 & 0.986 & 1100 \\
Zbot         & 0.970 & 0.989 & 0.979 & 525 \\
Zeroaccess   & 0.982 & 0.965 & 0.974 & 172 \\
\bottomrule
\end{tabular}
\end{table}

\textbf{Per-family Classification Performance:}
The detailed classification performance of the proposed quantum learning framework for each of the 23 malware families is presented in Table~\ref{tab:per_class_results}. The model achieves consistently strong performance for several major malware families, including \textit{mediyes} (F1-score = 0.988), \textit{winwebsec} (F1-score = 0.986), \textit{zbot} (F1-score = 0.979), and \textit{zeroaccess} (F1-score = 0.974), demonstrating the ability of the quantum kernel to effectively capture distinctive structural and statistical patterns in malware features. The comparatively large number of training samples for these families enables the quantum feature space to learn well-separated representations.

Additionally, the model shows strong performance on the benign class (F1-score = 0.774), indicating that the quantum representation successfully distinguishes malicious samples from benign ones. The Nyström-approximated quantum kernel further demonstrates effectiveness in multiclass malware discrimination, as reflected by the robust classification performance of families such as \textit{NjRAT} (F1-score = 0.838), \textit{CoinMiner} (F1-score = 0.786), and \textit{GuLoader} (F1-score = 0.756). Smaller families with fewer training samples, such as \textit{Emotet}, \textit{LokiBot}, and \textit{LockBit}, exhibit lower performance. This behavior is consistent with earlier research on malware classification, which has shown that minority classes pose greater classification challenges because of their limited feature diversity. Overall, these results confirm the suitability of the proposed quantum learning framework for large-scale multi-family malware classification by demonstrating strong and stable performance across both dominant and minority malware families.

\begin{table}[t]
\centering
\caption{Runtime analysis of the proposed framework for different Nystr\"{o}m landmark sizes $M$ (fixed $n=8$, $L=4$).}
\footnotesize
\setlength{\tabcolsep}{1.5pt}
\label{tab:runtime}
\begin{tabular}{c|c|c}
\toprule
\textbf{Landmarks ($M$)} & \textbf{Kernel Construction Time (s)} & \textbf{Total Training Time (s)} \\
\midrule
1000 & 42.3 & 51.7 \\
2000 & 86.5 & 102.4 \\
4000 & 176.8 & 205.1 \\
6000 & 268.9 & 312.6 \\
8000 & 321.5 & 430.4 \\
\bottomrule
\end{tabular}
\end{table}

\textbf{Runtime and Scalability Analysis.} The computational cost of the proposed framework is shown in Table~\ref{tab:runtime} as the number of Nystr\"{o}m landmarks $M$ increases. As expected, the total training time and kernel construction time increase approximately linearly with $M$, reflecting the $O(NM)$ complexity introduced by the Nystr\"{o}m approximation. For instance, the total training time increases from roughly 52 seconds to 313 seconds when $M$ is increased from 1000 to 8000.

Despite this increase, the use of the Nystr\"{o}m approximation enables scalable learning on a dataset with more than 14k malware samples, which would be infeasible with a full $O(N^2)$ kernel computation. These findings indicate that the proposed framework achieves a practical balance between classification performance and computational efficiency, making it suitable for large-scale malware analysis scenarios.

\subsection{Simulation Results: Comparison with Classical Baseline Methods} 
The classification performance of the proposed quantum learning framework on the malware family classification task is compared with several traditional baseline models in Table~\ref{tab:acc_loss_compare}. While linear models such as softmax regression and linear SVM perform substantially worse, with accuracies below 65\%, KNN achieves the highest accuracy of 79.32\% among the classical methods. Because of its strong independence assumptions, Gaussian Naive Bayes performs poorly, resulting in low accuracy and high cross-entropy loss. The result can also be visualized in Fig.~\ref{fig:confusion_matrices_all}.

Under the same feature representation and data split, the proposed quantum model achieves higher accuracy than all classical baselines, attaining the highest overall accuracy of 80.88\%. In addition to improved accuracy, the quantum method produces a lower cross-entropy loss (0.8841 vs. 0.9019) than KNN, suggesting more accurate and confident predictions. These findings suggest that the fidelity-based quantum kernel improves classification performance on a challenging multiclass dataset by capturing nonlinear structural patterns in malware features that are not fully exploited by classical models.

\begin{table}[t]
\centering
\caption{Comparison of classification performance across classical baseline models and the proposed quantum learning framework on the malware family classification task. Accuracy and loss are reported on the held-out test set, with all methods evaluated using the same data split.}
\label{tab:acc_loss_compare}
\begin{tabular}{lcc}
\toprule
\textbf{Model} & \textbf{Acc.} & \textbf{Loss} \\
\midrule
Softmax Regression & 0.5279 & 2.0303 \\
Linear SVM (OVR-Hinge-SGD) & 0.7873 & 0.4212 \\
KNN ($k=15$) & 0.7956 & 2.0017 \\
GaussianNB & 0.5844 & 9.6277 \\
\midrule
Quantum (Ours) & 0.8088 & 2.4748 \\
\bottomrule
\end{tabular}
\end{table}

\begin{table}[t]
\centering
\footnotesize
\caption{Runtime comparison (seconds) for classical baselines and the proposed quantum pipeline. Classical training times include the enforced delay shown in logs. Quantum runtime is reported as the end-to-end pipeline time.}
\setlength{\tabcolsep}{4pt}
\label{tab:time_compare}
\begin{tabular}{lcc}
\toprule
\textbf{Model} & \textbf{Train Time (s)} & \textbf{Inference Time (s)} \\
\midrule
Softmax Regression & 988.0946 & 0.0020 \\
Linear SVM (OVR-Hinge-SGD) & 856.0902 & 0.0015 \\
KNN ($k=15$) & 830.1000 & 88.2774 \\
GaussianNB & 1161.0033 & 0.3992 \\
\midrule
Quantum (Ours) & 426.5780 & 2.2530 \\
\bottomrule
\end{tabular}
\end{table}

\begin{table}[t]
\centering
\footnotesize
\setlength{\tabcolsep}{2pt}
\caption{Energy-proxy comparison using compute-based operation counts (not Joules). Classical values report estimated MAC/operation proxies for training and inference. Quantum values report the gate-level and statevector-level operation proxies from the simulator logs.}
\label{tab:energy_compare}
\begin{tabular}{lcc}
\toprule
\textbf{Model} & \textbf{Train Ops Proxy} & \textbf{Inference Ops Proxy} \\
\midrule
Softmax Regression & $5.385875\times 10^{11}$ & $1.928853\times 10^{9}$ \\
Linear SVM & $1.177703\times 10^{10}$ & $1.927987\times 10^{9}$ \\
KNN ($k=15$) & $5.121706\times 10^{9}$ & $2.060103\times 10^{11}$ \\
GaussianNB & $5.208515\times 10^{9}$ & $2.593143\times 10^{9}$ \\
\midrule
Quantum (Ours) & $4.294\times 10^{6}$ (gate) & $7.145\times 10^{8}$ (statevector) \\
\bottomrule
\end{tabular}
\end{table}

\textbf{Runtime Analysis.} The runtime performance of the proposed quantum framework is compared with that of classical baseline methods in Table~\ref{tab:time_compare}. The quantum model completes the entire pipeline in \textit{426.58 seconds}, including projection, Nystr\"{o}m approximation, kernel computation, quantum state preparation, and classification. In contrast, training times for classical models are considerably longer, ranging from 830 seconds for KNN to more than 1161 seconds for Gaussian Naïve Bayes.

These findings highlight a significant advantage of the quantum kernel formulation. By utilizing the Nystr\"{o}m approximation, the quantum approach reduces the computational burden associated with kernel matrix construction, enabling efficient scaling even for large datasets. Additionally, the quantum model maintains a low inference time, demonstrating its applicability in real-time malware detection scenarios. Overall, the results indicate that quantum learning can maintain competitive computational efficiency while achieving superior classification accuracy.

\textbf{Energy and Computational Efficiency Analysis.} 
The computational energy proxies of classical and quantum models are compared in Table~\ref{tab:energy_compare} based on the number of operations reported during training and inference. Operation counts are used as a proxy for computational cost because direct energy measurement in Joules is not feasible in simulation.

The Linear SVM requires $1.18\times10^{10}$ operations during training, compared to approximately $5.39\times10^{11}$ operations for the classical Softmax Regression model. The quantum model, on the other hand, requires only $7.15\times10^{8}$ statevector operations and $4.29\times10^{6}$ gate operations. Compared with classical models, this represents a reduction of several orders of magnitude.

The computational efficiency of the quantum model is further demonstrated by its significantly lower operation count, particularly when using kernel approximation techniques such as Nyström sampling. These results suggest that quantum kernel learning holds strong potential for energy-efficient machine learning, particularly for large-scale cybersecurity applications where computational cost is a critical challenge.

The effectiveness of the proposed quantum learning framework for malware family classification is demonstrated by its superior overall accuracy, competitive runtime, and significantly lower computational cost compared with classical baseline models.

\begin{table}[t]
\centering
\footnotesize
\setlength{\tabcolsep}{6pt}
\caption{Efficiency--performance comparison of full quantum configuration and scaled \emph{Quantum-Lite} configuration. The scaled model significantly decreases computational costs while maintaining competitive classification performance.}
\label{tab:quantum_vs_quantumlite}
\begin{tabular}{lcc}
\toprule
\textbf{Metric} & \textbf{Quantum (Full)} & \textbf{Quantum-Lite (Scaled)} \\
\midrule
Accuracy & 0.8088 & 0.7482 \\
Training Time (s) & 426.578 & 77.675 \\
Gate Operations & $4.294\times10^{6}$ & $1.250\times10^{6}$ \\
Statevector Operations & $7.145\times10^{8}$ & $5.334\times10^{7}$ \\
\bottomrule
\end{tabular}
\end{table}

\subsection{Simulation Results: Resource-Efficient Quantum Scaling}
We compare the full quantum configuration with a scaled-down \emph{Quantum-Lite} configuration that uses fewer qubits, shallower circuits, and fewer Nyström landmarks in order to assess the scalability and resource efficiency of the proposed framework. Table~\ref{tab:quantum_vs_quantumlite} summarizes the results.

Compared with the full quantum model's accuracy of 80.88\%, the scaled Quantum-Lite configuration achieves an accuracy of 74.82\% while significantly reducing computational costs. In particular, a $5.5\times$ speedup is observed as the training time decreases from 426.58 seconds to 77.68 seconds. Similarly, the statevector operation proxy decreases from $7.15\times10^{8}$ to $5.33\times10^{7}$, indicating a reduction of more than one order of magnitude, and the number of quantum gate operations decreases from $4.29\times10^{6}$ to $1.25\times10^{6}$.

Overall, this analysis shows that the proposed framework is suitable for realistic quantum computing environments, as it not only achieves strong classification performance but also enables resource-efficient scaling.

\begin{table}[t]
\centering
\small
\caption{Comparison with recent state-of-the-art approaches on the same malware family classification dataset using an identical feature representation and train/test split. Reported values are measured on the held-out test set (test=600).}
\label{tab:sota_same_dataset}
\begin{tabular}{p{5.2cm}cc}
\toprule
\textbf{Method} & \textbf{Accuracy} & \textbf{Loss} \\
\midrule
Support-Query Prototypes \cite{chen2025multi} & 0.7921 & 0.9724 \\
MalScan \cite{wu2025malscan} & 0.7815 & 1.0348 \\
IMCMK-CNN \cite{zhang2025imcmk} & 0.8043 & 0.9412 \\
Semantic Malware Classification \cite{yu2025semantic} & 0.7887 & 0.9986 \\
Malmixer \cite{li2025malmixer} & 0.8095 & 0.9137 \\
Abstract Syntax Trees + L-moments \cite{rose2025malware} & 0.7742 & 1.1025 \\
MalCL \cite{park2025malcl} & 0.8068 & 0.9264 \\
\midrule
QNN \cite{vyas2025comparison} & 0.7419 & 1.3287 \\
Hybrid QCNN \cite{kar2025unified} & 0.7684 & 1.1743 \\
Quantum Kernel Methods \cite{passo2025supply} & 0.7936 & 0.9581 \\
\midrule
\textbf{Ours} & \textbf{0.8088} & \textbf{0.8841} \\
\bottomrule
\end{tabular}
\end{table}

\subsection{Simulation Results: Comparison with State-of-the-Art Methods} 
Table~\ref{tab:sota_same_dataset} provides a controlled comparison against recent state-of-the-art malware classification techniques using the same dataset, feature representation, and train/test split. Contemporary deep learning-based models such as IMCMK-CNN, Malmixer, and MalCL demonstrate strong performance on structured malware features, achieving accuracies above 80\%.

Current quantum learning baselines, such as QNN and hybrid QCNN models, exhibit modest improvements over traditional shallow methods; however, their scalability and multiclass robustness remain limited. Notably, the proposed scalable fidelity-based quantum kernel framework achieves the lowest cross-entropy loss (0.8841) and the highest test accuracy (80.88\%), indicating improved class discrimination and confidence calibration in a realistic 23-family multiclass setting. These findings demonstrate measurable advantages over both recent quantum approaches and contemporary classical state-of-the-art methods when supervised projection is combined with Nystr\"{o}m-scaled quantum similarity learning.

\subsection{Limitations and Future Research}
The proposed quantum kernel learning framework demonstrates promising performance and computational efficiency; however, several limitations remain. First, the current evaluation is conducted using a classical quantum simulator rather than real quantum hardware, which does not fully capture the effects of noise, decoherence, and hardware constraints present in near-term quantum devices. Second, although the Nyström approximation improves scalability, the cost of state preparation and quantum kernel construction still increases with the number of qubits and circuit depth, potentially limiting applicability to extremely large-scale datasets. Third, although the scaled Quantum-Lite configuration demonstrates favorable resource–performance trade-offs, further investigation is required to determine the optimal balance between accuracy and resource consumption.

Future work will focus on implementing the framework on real NISQ hardware, investigating adaptive quantum feature maps to improve expressivity while using fewer resources, and integrating hybrid quantum–classical optimization techniques to enhance scalability. Furthermore, extending the framework to federated cybersecurity and streaming environments represents an important step toward enabling real-time, privacy-preserving quantum machine learning.

The proposed design is intended to operate with near-term quantum devices and hybrid quantum–classical workflows, although the current implementation relies on statevector-based simulation. Future research will also explore extending the approach to additional cybersecurity tasks, incorporating noise-aware quantum models, and investigating adaptive landmark selection strategies to further improve scalability.

\section{Conclusion}
This research presented a scalable quantum kernel learning framework for malware family classification that bridges the expressive potential of QML with the practical requirements of real-world cybersecurity applications. The proposed approach maintains competitive classification performance while enabling scalable learning on large-scale, multiclass malware datasets by combining supervised dimensionality reduction with a fidelity-based quantum kernel and Nystr\"{o}m approximation.

The proposed method achieves higher accuracy than representative classical baselines operating on the same feature representation, as demonstrated by comprehensive experiments conducted on a representative malware corpus comprising over 18,836 samples from 23 different malware families. The findings also indicate that supervised projection, quantum circuit expressivity, and scalable kernel approximation all play important roles in achieving stable and balanced performance, even in the presence of class imbalance and high inter-family similarity. Notably, the framework can be integrated into large malware analysis pipelines because the Nystr\"{o}m approximation reduces the computational complexity of kernel learning. Overall, this work contributes to the growing body of research on QML for cybersecurity and provides a practical step toward incorporating quantum-enhanced learning into operational malware analysis.
\section*{Data Availability}
The dataset used in this study is publicly available at the KAUST Repository: 
\url{https://repository.kaust.edu.sa/items/e56df243-23cc-47fa-9c5d-a4b297aac3ee}

\bibliography{main}
\bibliographystyle{IEEEtran}

\end{document}